\documentclass[12pt,preprint]{aastex}

\shorttitle{Electron Heating}
\begin{document}


\title{Black Hole Accretion in Low States: Electron Heating}


\author{Siming Liu,\altaffilmark{1} 
%Fulvio Melia,\altaffilmark{3} and 
Christopher L. Fryer,\altaffilmark{1, 2} and
Hui Li\altaffilmark{1}
%Vah\'e Petrosian,\altaffilmark{3} 
}

\altaffiltext{1}{Los Alamos National Laboratory, Los Alamos, NM 87545; liusm@lanl.gov; hli@lanl.gov}

%\altaffiltext{3}{Physics Department and Steward Observatory, The University of Arizona, 
%Tucson, AZ 85721; melia@physics.arizona.edu; Sir Thomas Lyle Fellow and Miegunyah Fellow.}
\altaffiltext{2}{Physics Department, The University of Arizona, Tucson, AZ 85721; clfreyer@lanl.gov}
%\altaffiltext{3}{Center for Space Science and Astrophysics, Department of Physics and Applied Physics, Stanford University, Stanford, CA 94305; vahe@astronomy.stanford.edu}


\begin{abstract}

Plasmas in an accretion flow are heated by MHD turbulence generated through the 
magneto-rotational instability. The viscous stress driving the accretion is intimately 
connected to the microscopic processes of turbulence dissipation. We show that, in a few 
well-observed black hole accretion systems, there is compelling observational evidence of 
efficient electron heating by turbulence or collective plasma effects in low accretion states, 
when Coulomb collisions are not efficient enough to establish a thermal equilibrium between 
electrons and ions at small radii. However, charged particles reach a thermal equilibrium with 
their kind much faster than with others through Coulomb collisions, a two-temperature 
accretion flow is expected. We consider a Keplerian accretion flow with a constant mass 
accretion rate in the pseudo-Newtonian gravitational potential and take into account the 
bremsstrahlung, synchrotron, and inverse Comptonization cooling processes. The critical mass 
accretion rate, below which the two-temperature solution may exist, is determined by the 
cooling processes and the collisional energy exchanges between electrons and ions and has very 
weak dependence on the collision-less heating of electrons by turbulence, which becomes more 
important at lower accretion rates. Collision-less heating of electrons by MHD turbulence can 
no longer be ignored in quantitative investigations of these systems.

\end{abstract}


%% subject header.  Objects should be in the appropriate "individual"
%% headers (e.g. quasars: individual, stars: individual, etc.) with the
%% additional provision that the total number of headers, including each
%% individual object, not exceed six.  The \objectname{} macro, and its
%% alias \object{}, is used to mark each object.  The macro takes the object
%% name as its primary argument.  This name will appear in the paper
%% and serve as the link's anchor in the electronic edition if the name
%% is recognized by the data centers.  The macro also takes an optional
%% argument in parentheses in cases where the data center identification
%% differs from what is to be printed in the paper.

%\object{NGC 6624}, \objectname[M 15]{NGC 7078},
%\object[Cl 1938-341]{Terzan 8})}


\section{Introduction}

Black hole accretion is one of the most powerful energy sources in the universe. When the luminosity of the 
system is close to the Eddington luminosity, the accretion can be described by the classical Shakura-Sunyaev 
disk (1973), which produces a multi-color blackbody radiation with the flux and temperature determined by the  
black hole mass and accretion rate. Observations of galactic black hole X-ray binaries strongly support such a scenario 
(Gierli\'{n}ski \& Done 2004). The less energetic non-thermal high energy emission component frequently observed has 
been attributed to hot magnetized coronas above the disk (Gierli\'{n}ski et al. 1999; Zhang et al. 2000). 

The magneto-rotational instability (MRI) has been generally accepted as the basic mechanism producing the less well-understood turbulence viscosity that drives the accretion (Balbus \& Hawley 1991). The dissipated gravitational energy is first converted into MHD turbulence, which then heats the accretion flow through viscous and Ohmic dissipations. 
In these optically thick slim disks, the Coulomb collision time scales are much shorter than other relevant time scales, electrons and ions reach a thermal equilibrium so that the observed emission can be well-described with blackbody spectra modified by the radiation transfer through the disk structure. The acceleration and/or heating of electrons by turbulence can at most be studied with observations of the non-thermal component emitted from the collision-less coronas (Dermer et al. 1996; Li \& Miller 1997). 

Many black hole candidates are often observed in a super- or sub- Eddington emission state, and there are many 
distinct observational phenomena, such as relativistic outflows and quasi-periodic oscillations, that have not 
been well understood though many models have been proposed (Fender et al. 2004; Remillard et al. 1999; Falcke 
et al. 2004; Camenzind 2005). Both theoretical investigations and observations suggest that, below a critical 
mass accretion rate $\dot{M}_{\rm cr}$, a two-temperature accretion flow likely develop near the black hole due 
to the much higher radiation efficiency of electrons than ions and inefficient Coulomb coupling between them so that the local thermal equilibrium can not be established between the two (Shapiro et al. 1976; Rees et al. 1982; Zdziarski et al. 2002; Titarchuk \& Fiorito 2004). Such a two-temperature flow is also expected due to the fact that charged 
particles reach a thermal equilibrium with their kind much faster than with others through Coulomb collisions 
(Spitzer 1962). This paper studies the electron heating processes in these low states.

Besides energy exchanges through Coulomb collisions, charged particles can also be energized by plasma waves through collision-less processes. These energization processes will directly affect the characteristics of the observed emission and therefore play crucial roles in our study of these systems in the low states.
As pointed out by Bisnovatyi-Kogan and Lovelace (1997), the Ohmic heating of electrons through 
dissipation of MHD turbulence can be very efficient. Sharma et al. (2007) recently found that electrons might also be energized in the dynamo processes of magnetic field amplification through the MRI.
However, the complexity of processes in a turbulent plasma has made the collision-less electron heating by collective plasma effects a difficult problem (Rees et al. 1982). In these theoretical investigations, certain assumptions have to be made on the coupling between charged particles and the turbulent electromagnetic fields to derive some quantitative results, which are usually sensitive to the prior assumptions.

The collision-less electron heating by turbulent magnetic fields is often ignored in most phenomenological models (Shapiro et al. 1976). In the advection dominated accretion flow models, it is simply assumed that a small fraction of the energy dissipated through viscosity is converted into electrons (Esin et al. 1998; Yuan et al. 2006), suggesting that this is a trivial process. Alternatively, one-temperature models have been proposed for 
Sagittarius A*, the compact radio source associated with the low-luminosity supermassive black hole in the Galactic Center (Sch\"{o}del et al. 2002; Ghez et al. 2005), assuming that electrons and protons are coupled by turbulence effectively (Melia et al. 2000, 2001).

On the other hand, the collision-less electron heating processes may be constrained by observations of the relevant systems. Recent studies of flares from Sagittarius A* indicate that electrons can be heated efficiently by MHD 
turbulence and the distribution of relativistic electrons under the influence of a turbulent magnetic field can 
be approximated as relativistic Maxwellian (Baganoff et al. 2001; Genzel et al. 2003; Gillessen et al. 2006; 
Bittner et al. 2007). The same processes may also play important roles in the heating of electrons in the 
two-temperature accretion flows. In this paper, we consider a Keplerian accretion disk model and the dominant 
cooling processes in a fully ionized magnetized plasma. We show that Coulomb collisions with ions can not heat 
electrons efficiently and an extra electron heating process is required to explain observations of Sagittarius 
A* and the galactic X-ray binary Cygnus X-1 (Wilms et al. 2006). This is in contradiction with what is suggested in some of the 
previous studies.  Magnetic turbulence could play such a role, and we believe that the 
collision-less heating of electrons by MHD turbulence should not be ignored in any quantitative studies of these low accretion states. 

The critical mass accretion rate 
%$\dot{M}_{\rm cr}$, below which a two-temperature may develop at small radii, 
is determined by the cooling processes and collisional energy 
exchanges between electrons and ions. Although the collision-less turbulence heating can reduce $\dot{M}_{\rm cr}$ by increasing the radiation efficiency and cooling rate, we find that this effect is insignificant. More 
detailed investigations of the radiation spectrum and magneto-hydrodynamic processes are needed to study these plasma processes quantitatively.

The basic equations for the accretion disk are given in \S\ \ref{dyeq}. In \S\ \ref{heating}, we discuss the 
electron heating processes and show that observations of Sagittarius A* in the millimeter and sub-millimeter 
range are difficult to explain without introducing efficient electron heating by turbulence. The cooling 
processes are studied in \S \ref{cooling} and the model is applied to galactic X-ray binaries in \S\ 
\ref{binary}. In \S\ \ref{dis}, we draw conclusions and discuss the model limitations and possible improvements 
in the future.

\section{Basic Equations for Keplerian Flows in the Pseudo-Newtonian Potential}
\label{dyeq}

We consider a fully ionized hydrogen plasma. Then the gas pressure and the thermal energy density are given, 
respectively, by
\begin{eqnarray}
P&=& n k_{\rm B} (T_p+T_e)\,, \\
{\cal E}&=& n k_{\rm B} (1.5 T_p + \alpha T_e)\,,
\end{eqnarray}
where the gas density, the proton and electron temperatures are denoted by $n$, $T_p$ and $T_e$, respectively, 
$k_{\rm B}$ is the Boltzmann constant, and $\alpha = x[3K_3(x)+K_1(x)-4 K_2(x)]/4K_2(x)$ with $x=m_ec^2/k_{\rm 
B} T_e$, where $m_e$ and $c$ denote the electron mass and the speed of light, respectively, and $K_i$ refers to 
the $i$th order modified Bessel function. $K_i(x)\rightarrow2^{i-1} (i-1)!/x^i$ as $x\rightarrow 0$ and 
$K_i(x)\rightarrow (\pi/2x)^{1/2}\exp({-x})[1+(4i^2-1)/8x]$ as $x\rightarrow \infty$. 

In the pseudo-Newtonian gravitational potential (Paczy\'{n}sky \& Wiita 1980), the potential and Keplerian angular velocity 
are given, respectively, by
\begin{eqnarray}
\phi &=& -{GM\over r-r_{\rm S}}\,,\\
\Omega_{\rm K}&=& \left[{GM\over r(r-r_{\rm S})^2}\right]^{1/2}\,,
\end{eqnarray}
where the gravitational constant, the black hole mass and the radius are denoted by $G$, $M$ and $r$ 
respectively, and $r_{\rm S}=2GM/c^2$ is the Schwarzschild radius of the black hole. 

Due to the decoupling between electrons and ions, the radiation efficiencies of two-temperature accretion flows will not be as high as the optically thick slim disks. The accretion processes likely drive strong winds from the hot disk (Blandford and Begelman 1999), and there are currently no strong observational constraints on these processes. To simplify our model we consider the steady state properties of the accretion flow averaged in the vertical direction and assume a radius independent accretion rate, 
i.e., ignoring the effects of winds: 
\begin{equation}
\dot{M} = - 4 \pi r H v_r n (m_p+m_e)\,,
\end{equation}
where $v_r$ and $H$ are the radial velocity and the scale height of the accretion flow, respectively, and  
$m_p$ is the proton mass. The vertical structure of the disk can be very complicated as suggested by MHD simulations (Hawley \& Balbus 2002) and may be essential to explain the observed non-thermal high energy emission in low states (Dermer et al. 1996; Li \& Miller 1997). It, however, is not expected to introduce significant changes to our quantitative results below, which deal with the energetically dominant thermal emission component from the main body of the hot accretion flow (Shapiro et al. 1976). 

Following Blandford and Begelman (1999), we assume that the angular momentum flux through the disk is negligible, which is also expected for strongly magnetized disks  (Agol \& Krolik 2000). MHD simulations also show that the azimuthal velocity of the flow is given by the Keplerian velocity $v_{\rm K} = r\Omega_{\rm K}$ (Hawley \& Balbus 2002), which requires the inner boundary radius $r_i\ge 3\ r_{\rm S}$, the radius of the last stable orbit. From the angular momentum conservation of the accretion flow, we then have
\begin{equation}
v_r = \nu {{\rm d}\ln \Omega_{\rm K}\over {\rm d} r}
%\left[1-{r_i^2\Omega_{\rm K}(r_i)\over r^2 \Omega_{\rm K}}\right]^{-1}
= -\nu \left({1\over 2r}+{1\over r-r_{\rm S}}\right)  
%\left[1-\left({r_i^3 (r-r_{\rm S})^2\over r^3(r_i-r_{\rm S})^2}\right)^{1/2}\right]^{-1}
\end{equation}
where the kinematic viscosity 
\begin{equation}
\nu={2\beta_p\beta_\nu k_{\rm B} (T_p+T_e)(r-r_{\rm S})^2r^{1/2}\over (m_p+m_e)(3r-r_{\rm S})(GM)^{1/2}}\,,
\end{equation}
and in accord with Melia et al. (2001) $\beta_p= \langle B^2\rangle /8\pi P$ and $\beta_\nu$ is defined as the 
ratio of the total stress to the magnetic field energy density $\langle B^2/8\pi\rangle $.\footnote{Note that 
the viscous parameter in the classical disk model $\alpha = \beta_\nu\beta_p/(1+2\beta_p)$.} Then the radial 
velocity can be rewritten as
\begin{equation}
v_r = -{\beta_\nu\beta_pk_{\rm B}(T_p+T_e)(r-r_{\rm S})\over (m_p+m_e)(GMr)^{1/2}}\,.
\end{equation}
%where we have defined $f\equiv \left[1-{r_i^{3/2} (r-r_{\rm S})/r^{3/2}(r_i-r_{\rm S})}\right]$. 
Since the magnetic field is dominated by the toroidal component, one has
\begin{eqnarray}
H &=& \left[{r k_{\rm B}(T_p+T_e)(1+2\beta_p)\over GM(m_p+m_e)}\right]^{1/2}(r-r_{\rm S})\,, \\
n &=& {GM\dot{M} (m_p+m_e)^{1/2}\over 4\pi \beta_\nu\beta_p [k_{\rm 
B}(T_p+T_e)]^{3/2}(1+2\beta_p)^{1/2}r(r-r_{\rm S})^2}\,.
\end{eqnarray}

The energy conservation equation is given by
\begin{equation}
{{\rm d} \over Hr{\rm d} r}\left\{Hrv_r[P(1+2\beta_p)+{\cal E}+n(m_p+m_e)[\phi-0.5v_{\rm 
K}^2+0.5v_r^2]]\right\} = -\Lambda\,,
\label{energy}
\end{equation}
where $\Lambda$ is the radiative cooling rate, and we have ignored any energy fluxes carried away from the disk by winds and waves and taken into account the effects of magnetic fields properly (Melia et al. 2001). Note that the work done by the torque force is given by  $-{{\rm d} [Hrv_r n(m_p+m_e)v_{\rm K}^2]/{\rm d} r}$. 
Then we have 
\begin{equation}
{{\rm d} \over {\rm d} r} {\varepsilon}= -{\Lambda\over v_r n}\,,
\label{eqtp}
\end{equation}
where
\begin{equation}
{\varepsilon} = k_{\rm B}[T_e(\alpha+1+2\beta_p)+T_p(2.5+2\beta_p)]+(m_p+m_e)[\phi-0.5v_{\rm K}^2+0.5v_r^2]\,
\end{equation}
can be considered as the energy of the accretion flow per proton and $\dot{M}\varepsilon/m_p$ gives the 
outwardly directed energy flux through the accretion disk (Blandford \& Begelman 1999). 

\section{Electron Heating}
\label{heating}

Because the viscous stress driving the accretion is induced by the MRI (Balbus \& 
Hawley 1991), the gravitational energy dissipation first generates MHD turbulence, which then heats the gas as the turbulence cascades from large scales to small scales. A fraction of the turbulence energy may also be carried away from the disk by MHD waves propagating toward large radii. However, this process is not expected to dominate the dynamics of the accretion flow at least for the relative more powerful states, when the hard X-ray luminosity of X-ray binaries can be a significant fraction of the Eddington luminosity. Very high accretion rates will be needed to produce the observed X-ray power if waves carry most of the dissipated gravitational energy toward large radii. The turbulence cascade also needs to be suppressed dramatically to make the wave escape process dominant. Therefore, in the steady state, the viscous heating rate 
$$\Gamma=v_r n (m_p+m_e) {{\rm d} [-\phi + 0.5 v_{\rm K}^2]\over{\rm d} r}$$ 
should be slightly greater than the turbulence cascade rate
$$\Gamma_B={C_2 c_SB^2\over 8\pi H},$$ 
where $C_2\sim1$ is dimensionless and determined by the properties of the turbulence and $$c_S = \left[{k_{\rm B}(T_i+T_e)(1+2\beta_p)\over m_p+m_e}\right]^{1/2}$$ is the speed of fast mode waves (Liu et al. 2006b). 
There is therefore an intimate connection between the viscous stress and turbulence dissipation.  
%We will show below $\beta_\nu= 2C_2/3$, i.e. the viscous stress needs to be comparable to the magnetic field energy density.
It is, however, not clear how this energy is distributed between electrons and protons in a collision-less 
plasma and what determines this energy partition, especially for hot plasmas in the two-temperature accretion 
flow of black holes. In the following, we will show how observations of low states of black hole accretion systems may be used to constrain the collision-less electron heating rate by turbulence.

The disk structure can be obtained once one specifies the electron heating and cooling rates and the electron and proton temperatures at the outer boundary. Following Blandford  and Eichler (1987), we have the electron heating time by turbulence (Liu et al. 2006b)
\begin{equation}
\tau_{\rm ac} = {3C_1 H \langle v_e\rangle \over c_S^2}\,,
\end{equation}
where $C_1$ is a dimensionless constant, and $$\langle v_e\rangle={2c (x+1)\over x^2 K_2(x)\exp(x)}$$ is the mean 
electron speed. Note that $\langle v_e\rangle /c\rightarrow (8k_{\rm B}T_e/\pi m_ec^2)^{1/2}$ as $x\rightarrow 
\infty$ and the electron heating becomes relatively more efficient at smaller radii if electrons become 
relativistic and $\langle v_e\rangle $ approaches to $c$. Then we have
\begin{equation}
{{\rm d} T_e\over {\rm d}r} = {T_e\over \tau_{\rm ac} v_r} +%{T_e\over n}{{\rm d} n\over {\rm d} r}+
{T_p-T_e \over \tau_{\rm Coul} v_r} -{\Lambda \over n \alpha k_{\rm B}v_r}\,,
\label{eqte}
\end{equation}
where 
$$\tau_{\rm Coul} = {3\pi m_em_p\langle v_e\rangle ^3\over 256 n e^4\ln \lambda}\,,$$
is the electron-proton energy exchange time through Coulomb collisions, $e$ is the elemental charge unit, and $\ln\lambda\simeq15$ for most 
astrophysical situations (Spitzer 1962). 
%The second term on the right hand side gives the heating due to compression.

There are several ways to estimate $C_1$. For a non-radiative accretion flow, $\Lambda=0$. At large radii, we 
can ignore the kinetic energy associated with the radial motion and $\alpha = 1.5$. If we ignore the Coulomb 
coupling term, we have the following solution:
\begin{eqnarray}
c_S&=& (k_{\rm B}(T_e+T_p)(1+2\beta_p)/(m_p+m_e))^{1/2} = [3(1+2\beta_p)/(5+4\beta_p)]^{1/2} v_{\rm K}\,,
\label{anasol1} \\
H    &=& [3(1+2\beta_p)/(5+4\beta_p)]^{1/2}r\,, \\
v_r &=& -[3\beta_\nu\beta_p/(5+4\beta_p)]v_{\rm K}\,, \\
n &=& [3/(5+4\beta_p)]^{-3/2}(1+2\beta_p)^{-1/2} \dot{M}/4\pi\beta_\nu\beta_p(m_p+m_e) r^2 v_{\rm K}\,,\\
k_{\rm B} T_e &=& [3/(5+4\beta_p)(1+2\beta_p)]^{-1} [\pi/72 C_1^2\beta_\nu^2\beta_p^2] m_e v_{\rm k}^2\,,
\label{eT} 
\\
T_p/T_e &=&  [3/(5+4\beta_p)]^{2} [72 C_1^2\beta_\nu^2\beta_p^2/\pi(1+2\beta_p)] (m_p+m_e)/m_e -1\nonumber \\
&=& 1.5\times 10^4 [C_1\beta_\nu\beta_p/(1+0.8\beta_p)]^2/(1+2\beta_p)-1
\label{anasol6}\,.
\end{eqnarray}
Then we have 
$$C_1 = 8.1\times 10^{-3} [(T_p/T_e+1)(1+2\beta_p)]^{1/2} (1+0.8 \beta_p)/\beta_\nu\beta_p>8.1\times 10^{-3} (1+2\beta_p)^{1/2}(1+0.8 
\beta_p)/\beta_\nu\beta_p\,.$$ It is interesting to note that the proton temperature satisfies an equation  
similar to equation (\ref{eqte})\footnote{With the assumptions adopted above, this equation actually can be 
derived from equations (\ref{energy}) and (\ref{eqte}). It is therefore not an independent equation.}:
$$
{{\rm d} T_p\over {\rm d}r} = {T_p\over \tau_{\rm ac} v_r}\,,
$$
suggesting that the heating time of protons is the same as electrons. Since the mean speed of protons $\langle 
v_p \rangle$ is usually different from $\langle v_e \rangle$, it is likely that the dimensionless constant 
$C_1$, which is determined by the microscopic physics of resonant and/or non-resonant coupling between particles and turbulence, is different for protons and electrons. Otherwise, $T_p/T_e=m_p/m_e$, and we have
\begin{equation}
C_1=[(5+4\beta_p)/18\beta_\nu\beta_p][\pi(1+2\beta_p)/2]^{1/2}\,.
\label{eqep}
\end{equation}

For this solution, the Coulomb energy exchanges between electrons and protons (with $\tau_{\rm Coul}$ independent of the radius $r$) become relatively more important 
at larger radii. So we have a one-temperature flow at large radii, which develops into the above solution at 
the radius, where the Coulomb energy exchange time becomes comparable to the accretion time:
\begin{equation}
\tau_{\rm vis} \equiv -r/v_r \simeq {(5+4\beta_p) r^{3/2} \over 3 \beta_\nu \beta_p (GM)^{1/2}}\,.
\label{acct1}
\end{equation}
The transition radius is therefore given by
\begin{eqnarray}
r_t/r_{\rm S} &=& \left({\pi^2 m_e m_p(m_p+m_e) GM c^3[1+2\beta_p]^2\over 3[5+4\beta_p] 2^{15/2} e^4 \ln 
\lambda 
C_1^3 \beta_\nu \beta_p \dot{M}}\right)^{2/3}
= \left({\pi^2 m_p [1+2\beta_p]^2\over 45[1+0.8\beta_p] 2^{13/2} m_e \ln \lambda C_1^3 \beta_\nu 
\beta_p}{L_{\rm Edd}\over \dot{M} c^2}\right)^{2/3}  \nonumber \\ 
&\simeq& \left({0.3[1+2\beta_p]^2 \over 
[1+0.8\beta_p] C_1^3 \beta_\nu \beta_p}{L_{\rm Edd}\over \dot{M} c^2}\right)^{2/3} \simeq 3.4\times 10^3 
\left({\beta_\nu^4\beta_p^4[1+2\beta_p]\over[1+0.8\beta_p]^8}\right)^{1/3} \left({L_{\rm Edd}\over \dot{M} 
c^2}\right)^{2/3}\,,
\label{transr}
\end{eqnarray} 
where $L_{\rm Edd} = 4\pi GM(m_p+m_e)c/\sigma_T$ is the Eddington Luminosity and $\sigma_T = (8\pi/3) 
e^4/m_e^2c^4$ is the Thomson scattering cross-section, and we have used the above solution to obtain the 
electron temperature and assumed $T_e=T_p$ to get the last expression. We see that two-temperature flows can readily develop at small radii even without considering the electron cooling processes. 

When Coulomb collisions become less efficient at smaller radii, in principle the distributions of electrons 
and protons are not necessarily Maxwellian. However, at a given mean energy, the time scales for protons and electrons reaching equilibrium with themselves are, respectively, $\sim (m_e/m_p)^{1/2}\tau_{\rm Coul}$ and $\sim (m_e/m_p)\tau_{\rm Coul}$, which are much shorter than $\tau_{\rm Coul}$, the time scale for electrons and protons reaching thermal equilibrium with each other (Spitzer 1962). The electron and proton distributions therefore may start to deviate from Maxwellian at $(m_e/m_p)^{2/3}r_t$ and $(m_e/m_p)^{1/3} r_t$, respectively. So the 
transition radius, where the proton distribution may deviate from Maxwellian, is more than ten times smaller 
than the radius, where the two-temperature flow starts to develop. The transition radius for electrons is even
smaller, and a Maxwellian distribution can be achieved at small radii through relativistic effects and cooling processes (Wolfe \& Melia 2006; Bittner et al. 2007). We therefore expect that the electron distribution be always dominated by a thermal component and, over a significant radius range, a two-temperature flow develop self-consistently. The fact that the low-hard state spectra of X-ray binaries can be fitted with the thermal inverse Comptonization model also suggests a dominant thermal electron component. 

At very low accretion rates, the Coulomb collision time scales can be much longer than other relevant time scales near the black hole, and we do not expect Maxwellian distributions for both electrons and ions. This is a completely new regime, where the behavior of the turbulent plasma has not been well studied. In these cases, the ``temperature'' of the particles should be interpreted as a measurement of their mean energies. The actual electron distribution may be constrained by studying its radiation spectrum. We note that the proton distribution is always expected to deviate from Maxwellian at very small radii, where the energy dissipation rate is high, and a high energy proton population may be responsible for the formation of jets in the low states (Dermer et al. 1996; Liu et al. 2006a). A detailed investigation of these effects is beyond the scope of this paper.

From $\Gamma>\Gamma_B$ and the above solution, we have $\beta_\nu > 2C_2/3$. The viscous stress therefore needs to be comparable to the turbulent magnetic field energy density. Since the viscous stress should not be much greater than the turbulent magnetic field energy density (Pessah et al. 2006), the fact that $C_2\sim1$ requires that $\Gamma_B$ can not be much less than $\Gamma$. So the energy carried away by winds and/or waves should account for a small fraction of the dissipated gravitational energy except that the turbulence cascade from large scales to small scales is suppressed for some reason so that $C_2$ becomes much less than $1$.

We also note that for this solution, $\tau_{\rm ac}\propto r^{3/2}$. The turbulence electron heating effectively converts a constant fraction 
$\delta$ of the viscously dissipated gravitational energy into the internal energy of electrons. The electron heating rate by turbulence is given by $\alpha n k_{\rm B} T_e/\tau_{\rm ac}$. Then $$\delta\simeq [\pi 
(5+4\beta_p)(1+2\beta_p)/216C_1^2\beta_\nu^2\beta_p^2]m_e/(m_p+m_e)<1$$ gives $C_1\beta_\nu\beta_p>6.3 \times 
10^{-3} [(1+0.8\beta_p)(1+2\beta_p)]^{1/2}$. Although this result is similar to that proposed in the advection 
dominated accretion flow models (Esin et al. 1998), the electron heating studied here becomes very efficient when cooling reduces the 
electron temperature or the electrons become relativistic since the heating time is proportional to the mean 
velocity of electrons. Accordingly, Coulomb collisional heating may suppress the turbulence heating by 
increasing $\langle v_e\rangle$. 

Since the heating is more efficient when electrons become relativistic, an even stricter constraint on $C_1$ 
can be obtained by considering the accretion processes at small radii. For $\Lambda = 0$ and $T_e\gg T_p$, 
$\alpha = 3$, the accretion time is given by
\begin{equation}
\tau_{\rm vis} \equiv -r/v_r \simeq {(8+4\beta_p) r^{3//2} \over 3 \beta_\nu \beta_p (GM)^{1/2}}\,.
\end{equation}
To avoid electrons being overheated, the electron heating time needs to be longer than the accretion time, then 
we have:
\begin{equation}
C_1\beta_\nu\beta_p>\left({8+4\beta_p\over 27}\right)^{1/2}\left({GM\over r c^2}\right)^{1/2} < 0.3\,,
\end{equation}
where we have used the fact that, for the accretion flow we are considering, $r>3 r_{\rm S}$.

For $\Lambda=0$, we solve equations (\ref{eqtp}) and (\ref{eqte}) numerically to 
constrain the heating rate more quantitatively. 
%To satisfy the zero stress condition at the inner boundary $r_i = 3 r_{\rm S}$, 
%the radial velocity $v_r$ must diverge there and the radial motion can carry a huge amount of kinetic energy. 
%In reality, magnetic fields can produce significant stress at $r_i$ and $v_r$ never diverges. We will ignore 
%this degree of freedom in our modeling and only consider the range where $r>4r_{\rm S}$. As far as 
For $\beta_p=1$,  we find that $C_1\beta_\nu\beta_p$ needs to be greater than $0.15$ to avoid the proton temperature decreasing below zero at small radii at certain accretion rates due to efficient electron heating. Panel ``{\rm a}'' of Figure \ref{f1.eps} shows the temperature profiles for $\dot{M} = 10^{-7}, 10^{-3}, 10 L_{\rm 
Edd}/c^2$. (Note that the Eddington accretion rate is usually defined as $\dot{M}_{\rm Edd}\equiv 10 L_{\rm Edd}/c^2$.) The other model parameters $C_1=1.5$, $\beta_\nu = 0.1$, $\beta_p = 1$, and $T_e=T_p = GMm_p/10 k_B r_o$ at the outer boundary $r_o = 10^4 r_S$. When $\dot{M}$ is small, Coulomb collision effects are negligible. We recover the solution given by equations 
(\ref{anasol1}---\ref{anasol6}) at large radii. At small radii, the electron temperature goes beyond that given 
by equation (\ref{eT}) due to the relativistic effect discussed above. With the increase of $\dot{M}$, the 
electron and proton temperatures become closer to each other at larger radii. At small radii, because of 
efficient heating of relativistic electrons, the electron temperature is higher than the proton temperature for 
$\dot{M} = 10^{-3} L_{\rm Edd}/c^2$. When the accretion rate becomes comparable to or higher than the Eddington accretion rate $\sim10 L_{\rm Edd}/c^2$, the electron and proton temperatures become identical due to very efficient Coulomb collisions.

The linearly polarized millimeter and sub-millimeter emission from Sagittarius A* (Aitken et al. 2000; Sch\"{o}del et al. 2002; Ghez et al. 2005) reveals a mass 
accretion rate below $10^{-5}L_{\rm Edd}/c^2$ (Quataert \& Gruzinov 2000). X-ray observations and detailed modeling suggest an even lower accretion rate (Baganoff et al. 2001; 
Melia et al. 2000; Liu \& Melia 2002). VLBI observations have shown that this emission comes from a region 
within $\sim15\ r_{\rm S}$ of the black hole and should originate from a hot accretion torus (Shen 2005; Melia et al.
2001). The dotted line in Panel ``{\rm a}'' of Figure \ref{f1.eps} shows the electron (lower) and proton 
(upper) temperature profiles for $\dot{M} = 10^{-5} L_{\rm Edd}/c^2$ and $C_1=15$, which is ten times bigger than that for the other lines, giving much 
less efficient electron heating by turbulence. Although we have adopted a relatively high accretion rate, Coulomb collisions are 
not efficient enough to heat electrons to relativistic energies required to produce the observed synchrotron 
emission. A fit to the millimeter and sub-millimeter spectrum and polarization gives $C_1\beta_\nu\beta_p\simeq 
0.1$, which is comparable to the maximum heating rate discussed above, where cooling is absent, and is more than 10 times smaller than that given by equation (\ref{eqep}) (Liu et al. 2007). One can show that $C_1$ for electrons is more than 10 times smaller than 
that for protons in this case. This could be due to the high mean momentum of protons so that the action of protons on 
turbulence in their stochastic scattering processes is important. The proton heating is then suppressed giving 
rise to a more than $10$ times longer acceleration time than electrons at a given mean speed of both particles.

\section{Electron Cooling}
\label{cooling}

For a fully ionized magnetized plasma, synchrotron, inverse Comptonization (IC), and bremsstrahlung are the 
dominated emission mechanisms (Rybicki \& Lightman 1979). In the low accretion states of black holes, the disk 
is optically thin to bremsstrahlung radiation. We therefore have the bremsstrahlung cooling rate
\begin{equation}
\Lambda_{\rm brem} = \left({2\pi k_{\rm B} T_e\over 3 m_e}\right)^{1/2} {32\pi e^6\over 3 h m_e c^3} n^2 
g_{\rm B} =1.4\times 10^{-27} T_e^{1/2} n^2 g_{\rm B}\,, 
\end{equation}
where $g_{\rm B}\simeq 1.2$ is the Gaunt factor and $h$ is the Planck constant. For equations 
(\ref{anasol1})-(\ref{anasol6}), $\Lambda_{\rm brem}\propto r^{-7/2}$ and the bremsstrahlung cooling time of 
electrons $\tau_{\rm bc}= \alpha k_{\rm B} nT_e/\Lambda_{\rm brem}\propto r$.   Including this energy loss 
process, we find that the disk collapses to a low temperature flow toward small radii when 
$\dot{M}>\dot{M}_{\rm cr}=2.63\times 10^{-2} L_{\rm Edd}/c^2$ for an outer boundary radius of $10^4 r_{\rm S}$. 
This can be understood by comparing $\tau_{\rm bc}$  with the accretion time $\tau_{\rm vis}\propto r^{3/2}$ 
[See eq. (\ref{acct1})]. Then we have
\begin{equation}
\dot{M}_{\rm cr} = {3\sqrt{3m_p/\pi m_e} h \beta_\nu^2\beta_p^2v_{\rm K}\over 8 e^2 g_{\rm B}}\left({3\over 
5+4\beta_p}\right)^3{L_{\rm Edd}\over c^2} \simeq 2.9 (r_{\rm S}/r)^{1/2} {L_{\rm Edd}\over c^2}\,,
\label{mdotbrem}
\end{equation}
where we have assumed that $T_e=T_p$, i.e. $C_1 = [(5+4\beta_p)/18\beta_\nu\beta_p][\pi(1+2\beta_p)m_e/(m_e+m_p)]^{1/2}$, and used the model parameters for Panel ``{\rm a}'' of Figure 
\ref{f1.eps} to obtain the last expression. The critical mass accretion rate decreases with the increase of 
the outer boundary radius. 

Panel ``{\rm b}'' of Figure \ref{f1.eps} shows the temperature profiles for several 
values of $\dot{M}$. The other model parameters are the same as Panel ``{\rm a}''. Note that the 
bremsstrahlung cooling has little effect on the temperature profile for $\dot{M}\ll\dot{M}_{\rm cr}$ because, 
compared with the accretion time given by equation (\ref{acct1}), this cooling becomes less important toward smaller radii. For 
$\dot{M}=\dot{M}_{\rm cr}$, we see that the temperatures are lowered by the cooling near the outer boundary. 
Since the cooling rate is proportional to the density square, the cooling time scale is proportional to $M$,  
and $\dot{M}_{\rm cr}c^2/L_{\rm edd}$ is independent of $M$. Pair production will increase the cooling rate  and decrease $\dot{M}_{\rm cr}$. These results are similar to those given by Narayan \& Yi (1995) though electron heating by turbulence has been included in our calculations, implying that the collision-less electron heating does not affect the bremsstrahlung cooling rate at large radii significantly.

One may define a critical radius with equation (\ref{mdotbrem}):
\begin{equation}
r_{\rm cr}/r_S = {27 m_p h^2 \beta_\nu^4\beta_p^4c^2\over 128 \pi m_e e^4 g^2_{\rm B}}\left({3\over 
5+4\beta_p}\right)^6\left({L_{\rm Edd}\over \dot{M} c^2}\right)^2\simeq 8.4\left({L_{\rm Edd}\over \dot{M} c^2}\right)^2\,.
\end{equation}
A hot accretion flow can only exist below $r_{\rm cr}$. When the transition radius $r_t$ given by equation (\ref{transr}) is greater than $r_{\rm cr}$, as is the case for relatively high accretion rates, a hot two-temperature flow may develop below $r_{\rm cr}$. For low accretion rates, $r_{\rm cr}$ can be much greater than $r_t$, we have a hot one-temperature flow between $r_t$ and $r_{\rm cr}$, which develops into a hot two-temperature flow below $r_t$. The thin lines in Panel ``b'' corresponds to this case, where $r_t\sim 10^3 r_S$ and $r_{\rm cr} = 10^4 r_S$. From $r_t=r_{\rm cr}$, we have $\dot{M} c^2/L_{\rm Edd} \simeq 156 \beta_\nu^2\beta_p^2/(1+2\beta_p)^{1/4}(1+0.8\beta_p)^{5/2}$ and $r_t=r_{\rm cr}\simeq117 r_S (1+2\beta_p)^{1/2}/(1+0.8\beta_p)$.
However, how exactly the hot flow may develop toward small radii is still an open question, which may depend on the nature of the large-scale flow and the coupling between the disk and its corona (Meyer-Hofmeister et al. 2005).

When the source is optically thin to synchrotron radiation, the cooling rate is given by
\begin{equation}
\Lambda_{\rm syn} = {4e^4n\over 9 m_e^4 c^5} \langle p^2\rangle B^2 \simeq 1.06\times 10^{-15}n B^2 
{3x^2+12x+12\over x^3+x^2}\,,
\label{Lsyn1}
\end{equation}
where $\langle p^2\rangle $ is the mean momentum square of the electrons (Melia \& Coke 1999).
For equations (\ref{anasol1})-(\ref{anasol6}) and $x\ll1$, $\Lambda_{\rm syn}\propto r^{-6}$. The 
corresponding cooling time scale $\tau_{\rm sc} \propto r^{7/2}$, which is more important at smaller radii.  
For $x\gg1$, $\Lambda_{\rm syn}\propto r^{-5}$ and $\tau_{\rm sc}\propto r^{5/2}$. However, most of the 
thermal synchrotron emission is emitted at 
$$\nu_E \simeq {x+20\over x+1} \nu_c = {60(1+0.05x) eB\gamma_c^2 \over 4\pi(x+1) m_e c}=8.4\times 10^7 
B\gamma_c^2(1+0.05x)/(x+1)\, {\rm Hz}\,,$$
where $\gamma_c = 1/x+1$ (Liu et al. 2006). The thermal synchrotron emission (emissivity per solid angle) and 
absorption coefficients are given, respectively, by (Petrosian 1981; Mahadevan, Narayan, \& Yi 1996)
\begin{eqnarray}
{\cal E}_\nu &=&{\sqrt{3} e^3\langle p^2\rangle  (x+1)\over 8\pi m_e^3 c^4\langle \gamma^2\rangle (1+0.19x)} 
B\, n\, z_M\, I(z_M)  \,,\\
\kappa_\nu &=& {{\cal E}_\nu c^2 [\exp{(h\nu/k_{\rm B}T_e)}-1]\over 2 h \nu^3} \simeq {\pi e n\langle 
p^2\rangle \over 3\sqrt{3}\gamma_c^4m_e\langle \gamma^2\rangle Bk_{\rm B} T_e}{I(z_M)(x+1)\over z_M(1+0.19x)}
%=2.482\times 10^{-12} C_1^5 n_6^6 R_{13}^5 B_0^{-1} {I(z_M)\over z_M} {\rm cm^{-1}}
\,.
\end{eqnarray}
where $\langle \gamma^2\rangle \simeq (x+12)(x+1)/x^2$ is the mean Lorentz factor square of electrons, 
$(x+1)/(1+0.19x)$ is chosen to take into account the non-relativistic effects (see below), and
\begin{eqnarray}
I(z_M)&=& 4.0505z_M^{-1/6}(1+0.40z_M^{-1/4} + 0.5316z_M^{-1/2})\exp(-1.8899\,z_M^{1/3})\,,
\label{Im}
\\ 
z_M &=& {\nu/\nu_c} 
\equiv {4\pi m_e c\nu/3 e B \gamma_c^2} 
%= 14.12\ C_1^2\ \nu_{11} R_{13}^2\ n_6^2\ B_0^{-1}
\,.
\label{xm}
\end{eqnarray}
We then have the optical depth through the emission region 
\begin{eqnarray}
\tau_\nu(\nu) &\equiv& \kappa_\nu(\nu) H = {\pi e \langle p^2\rangle  n\ H\over 3\sqrt{3}m_e\langle 
\gamma^2\rangle k_{\rm B} T_e\gamma_c^4B} 
%= 24.82\ C_1^5\ R_{13}^6\ n_6^6\ B_0^{-1} [I(z_M)/ z_M] 
{I(z_M)(x+1)\over z_M(1+0.19x)}
%\equiv \tau_a {I(z_M)\over z_M}
%\nonumber \\
%&=& 
%1.783\times 10^{-7} 
%C_1^{-5} C_2^{-16/7} D_8^{-8/7}
%\left({x_{M}\over7.314}\right)^{-2}
%\left[{x_{M} I(x_{M})\over0.7811}\right]\,,
% \nonumber \\
%&& 
%\left({x_{M}\over7.314}\right)^{19/7}
%\left({x_{M} I(x_{M})\over0.7811}\right)^{4/7}
%\left({\nu_{14}\over 1.429}\right)^{-19/7} 
%\left({F_\nu\over 7\ {\rm mJy}}\right)^{-4/7} 
\,,
\end{eqnarray}
 which, for a given $z_M$, is proportional to $r^{23/4}$ for equations (\ref{anasol1})-(\ref{anasol6}) and 
$x\ll 1$ and proportional to $r^{3/4}$ for $x\gg1$. When $\tau_\nu(\nu_E)$ becomes greater than unity at large radii, one has to take into account the self-absorption effects. The corresponding synchrotron cooling rate may 
be approximated as
\begin{eqnarray}
\Lambda_{\rm syn}& = &{16\sqrt{3}\pi^2k_{\rm B} T_e\nu_E^3 \over 45^2 I[(x+20)/(x+1)] c^2H}{\langle 
\gamma^2\rangle (1+x)(1+0.19x)\over \gamma_c^2(1+0.05x)^2} \nonumber \\
&=&{80\sqrt{3} e^3 k_{\rm B} \over 3\pi I[(x+20)/(x+1)]m_e^3c^5} {T_e B^3\langle \gamma^2\rangle 
\gamma_c^4(1+0.05x)(1+0.19x)\over H(x+1)^2} \nonumber \\
%&\simeq& 6.4\times 10^{-13} {T_e B^3\langle \gamma^2\rangle \gamma_c^4(1+0.05x)(1+0.19x)I(20)\over H(x+1)^2I[(x+20)/(x+1)]} \nonumber \\
&\simeq& 6.4\times 10^{-13} {T_e B^3\over H}{(x+12)(x+1)^3(1+0.05x)(1+0.19x)I(20)\over x^6I[(x+20)/(x+1)]}\,,
\label{Lsyn2}
\end{eqnarray} 
where the numerical factor is chosen so that the cooling rates given by equations (\ref{Lsyn1}) and 
(\ref{Lsyn2}) are equal when $\tau_\nu(\nu_E) = 1$, and $\Lambda_{\rm syn}\simeq 8\pi k_{\rm B} T_e 
\nu_E^3/3c^2H$ for  $x\gg1$.  That is, for cyclotron radiation in the optically thick regime, we assume a 
black body spectrum cutting off at $\nu_E=\nu_c$. This cooling is important at small radii since $\Lambda_{\rm 
syn}\propto r^{-47/4}$ and $\tau_{\rm sc}\propto r^{37/4}$ for equations (\ref{anasol1})-(\ref{anasol6}) and 
$x\ll 1$. For $x\gg 1$, $\Lambda_{\rm syn}\propto r^{-23/4}$ and $\tau_{\rm sc}\propto r^{13/4}$.

We will assume that the IC is in the Thomson limit and the seed photons for IC are provided by the synchrotron 
radiation. Then we have the total cooling rate
\begin{eqnarray}
\Lambda &=& \Lambda_{\rm syn} + \Lambda_{\rm IC} + \Lambda_{\rm brem} 
= \Lambda_{\rm syn} [1+ 8\pi (\Lambda-\Lambda_{\rm brem})H/cB^2] +\Lambda_{\rm brem} \\
&=&\Lambda_{\rm syn}(1-8\pi\Lambda_{\rm syn}H/cB^2)^{-1} +\Lambda_{\rm brem} %\nonumber \\
%&=&{4e^4n\over 9 m_e^4 c^5} \langle p^2\rangle B^2(1-8\pi \Lambda_{\rm syn}H/cB^2)^{-1} +\left({2\pi k_{\rm B} Te\over 3 m_e}\right)^{1/2} {32\pi e^6\over 3 h m_e c^3} n^2 g_{\rm B} \nonumber \\
%&=&1.06\times 10^{-15}n B^2 {3x^2+12x+12\over x^3+x^2} (1-8\pi \Lambda_{\rm syn}H/cB^2)^{-1}  +1.4\times 10^{-27} T_e^{1/2} n^2 g_{\rm B}
\,. \nonumber
\end{eqnarray} 
%where $\Lambda_{\rm syn}, \Lambda_{\rm IC}$ and $\Lambda_{\rm brem}$ are the cooling rates due the synchrotron, IC and bremsstrahlung radiation, respectively, 
Panel ``{\rm c}'' of Figure \ref{f1.eps} shows the temperature profiles with the above cooling processes 
included for $M = 3.4\times 10^6M_\odot$. The model parameters are the same as Panel ``{\rm b}'' except 
$\dot{M} = 2.63\times 10^{-6}$ (thick), $2.63\times 10^{-4}$ (medium), $2.63\times 10^{-2}$ (thin) $L_{\rm 
Edd}/c^2$. We note that the electron temperature never goes above $\sim 10^{11}$ K, which may provide an 
alternative explanation to the observed upper limit in brightness temperature of powerful extragalactic radio 
sources (Readhead 1994). The critical mass accretion rate $\dot{M}_{\rm cr} = 2.63\times 10^{-2}L_{\rm 
Edd}/c^2$ is the same as that obtained above, where only the bremsstrahlung cooling is considered. This is 
consistent with the fact that the synchrotron and IC coolings are unimportant at large radii.  At small radii, 
these cooling processes do reduce the electron temperature. 
%For $\dot{M}\simeq\dot{M}_{\rm cr}$, the electron temperature saturates at a level determined by the balance between (optically thin) synchrotron cooling and Coulomb collisional heating:
%\begin{equation}
%\beta_p(1+2\beta_p) \pi^2 m_p (x+2)^2(x+1)^2=\alpha\ln\lambda m_e x^8 K^3_2(x)\exp(3x)\,,
%\end{equation}
%where we have assumed that $T_e\ll T_p$.

Because the synchrotron cooling rate in the optically thick regime is proportional to $n^{3/2}/H$, the cooling 
time is proportional to $M^{3/2}$. This cooling becomes relatively more efficient for smaller black holes so 
that the electron temperature decreases with the decrease of $M$ for given $\dot{M}c^2/L_{\rm Edd}$. %However, because the cooling is unimportant at large radii, $\dot{M}_{\rm cr}$ remains the same.
$\dot{M}_{\rm cr}$ also decreases slightly with the decrease of $M$. For $M = 3.4 M_\odot$, we find $\dot{M}_{\rm cr} = 2.13\times 10^{-2} L_{\rm Edd}/c^2$. 
Panel ``{\rm d}'' of Figure \ref{f1.eps} is the same as Panel ``{\rm c}'' except that $M=3.4 M_\odot$. 
%We also note that the radius (in units of $r_{\rm S}$) where the synchrotron cooling becomes optically thin, i.e. $\tau_\nu(\nu_E)=1$, increases with the decrease of $M$. 
The flatten part of the electron temperature in the inner region for $\dot{M} \sim \dot{M}_{\rm cr}$ is 
dominated by Coulomb collisional heating and optically thick synchrotron cooling, which is very sensitive to 
the electron temperature when $x<1$.  
%The increase of the synchrotron cooling efficiency for smaller black holes also results in lower electron temperatures. 
We therefore expect that, for $\dot{M} = \dot{M}_{\rm cr}$, the intrinsic brightness temperatures for 
extragalactic radio sources are higher for bigger black holes and there should be a correlation between the 
brightness temperature and the source luminosity. Pair production effects need to be incorporated to give more 
quantitative predictions. 

Due to the low radiation efficiency at large radii, the critical mass accretion rate 
is not very sensitive to the electron heating rate either.  Panel ``{\rm e}'' of Figure \ref{f1.eps} show the 
temperature profiles for $C_1 = 15$. The other parameters are the same as Panel ``{\rm d}'', and $\dot{M}_{\rm 
cr}=2.51\times 10^{-2}L_{\rm 
Edd}/c^2$. 

\section{Application to X-ray Binaries in the Low-Hard States}
\label{binary}

It has been suggested that Coulomb collisions are efficient enough to heat electrons to explain observations of galactic X-ray binaries in the low-hard states (Shapiro et al. 1976; Esin et al. 1998). Although some of the relevant models may achieve an electron temperature in line with observations by adjusting parameters describing the source structure and/or dissipation processes, the related assumptions  are not well justified both theoretically and observationally. It is also not obvious whether these models can explain the recently observed anti-correlation between the electron temperature and the Thomson optical depth of Cygnus X-1 over a large dynamical range (Wilms et al. 2006). The model proposed here is self-consistent and has the basic parameters to describe the dynamics and radiative processes. It is therefore well positioned to uncover new physical processes when applied to specific observations.

With the parameters adopted above, the model clearly can not explain the high hard X-ray luminosities observed in some of the galactic X-ray binaries.  The hard X-ray luminosity can be a significant fraction of $L_{\rm Edd}$ for some sources.  A much higher $\dot{M}_{\rm cr}$ is needed to produce such a high luminosity. To increase $\dot{M}_{\rm cr}$, one may increase the viscosity by increasing $\beta_\nu$ or decrease the outer 
boundary radius. The virial temperature at $r=10^4r_{\rm S}$ is already a few tens of keV, and we are 
interested in showing how electrons reach this temperature range from a large-scale cool accretion flow. We 
will consider the first option and set the outer boundary temperature $k_{\rm B}T_e=k_{\rm B}T_p=5\times 
10^{-6}m_pc^2\simeq 5$ keV in the following. Panel ``{\rm a}'' of Figure \ref{f2.eps} shows the temperature 
profiles for $M=34M_\odot$, $C_1=0.2$, $\beta_p = 1$, and $\beta_\nu = 3$. The critical mass accretion rate $\dot{M}_{\rm cr}= 21.9 L_{\rm Edd}/c^2$, which is consistent with equation (\ref{mdotbrem}). Therefore, if the hot accretion flow develops naturally with the processes considered here, a very high viscosity is needed to produce a hard X-ray luminosity on the order of $0.1 L_{\rm Edd}$. This statement is not necessarily true if the formation of hot accretion flows is triggered by some instability near black holes.

However, for $\dot{M}\sim \dot{M}_{\rm cr}$, the electron temperature is more than 1 MeV near the black hole, which is much higher than the observed values in the low-hard states (See Fig. \ref{f2.eps} a). It is clear that, besides the local internal cooling processes considered above, other cooling processes have to be introduced to bring the electron temperature to the desired energy range. Observations of X-ray binaries in the low-hard states show that photons from the large-scale optically thick slim disk can provide sufficient soft photons to cool electrons in the inner two-temperature flow through IC (Barrio et al. 2003). The cooling rate will also increase if we consider a global radiation transfer and include the cooling due to pair production. Here we treat these cooling processes approximately by assuming the energy density of the soft photons produced outside the two-temperature accretion flow is three times higher than the local magnetic field energy density so that there is an extra cooling term 3 times higher than that given by equation (\ref{Lsyn1}).
Panel ``{\rm b}'' of Figure \ref{f2.eps} shows the temperature profiles with this cooling effect 
included.  In this case, $\dot{M}_{\rm cr} = 
0.933L_{\rm Edd}/c^2$, which is clearly affected by this extra cooling process. The other model parameters are 
the same as Panel ``{\rm a}''.  

According to this model, the IC of the external soft photons dominates the 
cooling, and the electron temperature is nearly constant at small radii due to the balance between Coulomb collisional heating and this cooling for $\dot{M}=\dot{M}_{\rm cr}$. The value of the electron temperature depends on the ratio of the external photon 
energy density to that of the magnetic field $\eta$:
\begin{equation}
%3x^2+12x+12/(x+1)x^{7/2}
k_{\rm B} T_e \simeq \left({9\ln \lambda m_e\over 4\sqrt{2\pi} \beta_p \eta m_p}\right)^{2/5} m_e c^2\,.
\end{equation}
The electron heating by turbulence becomes relatively more important for lower values of $\dot{M}$. However, with 
$\dot{M}$, which is proportional to the Thomson 
optical depth of the accretion flow,
 varying by two orders of magnitude, the electron temperature in the inner region remains about $30$ 
keV. This is not consistent with the recently observed anti-correlation between the 
electron temperature and the Thomson optical depth of Cygnus X-1 in the low-hard and intermediate states (Wilms 
et al. 2006; Titarchuk 1994).

By increasing the electron heating rate by one order of magnitude, i.e. with $C_1 = 0.02$, we obtain temperature profiles in Panel ``{\rm c}''. Because IC cooling is already important at the outer boundary, $\dot{M}_{\rm cr} = 0.891 L_{\rm 
Edd}/c^2$ decreases slightly due to more efficient electron heating, which results in more cooling of the 
accretion flow. The other model parameters are the same as Panel ``{\rm b}''. We see that the electron 
temperatures are in the observed energy range. Panel ``{\rm d}'' shows profiles of the Thomson optical depth 
$\tau =2 \sigma_T H n$ and $\varepsilon$ for the same models in Panel ``{\rm c}''. We qualitatively recover the 
observed anti-correlation between $k_{\rm B} T_e$ and $\tau$. The radiation efficiency of the accretion flow is 
$\sim 0.01$ giving rise to a luminosity of $\sim 10^{35-38}$ erg s$^{-1}$, which is also in line with 
observations. We note that $C_1\beta_\nu\beta_p= 0.06$ implying very efficient electron heating by turbulence.

\section{Conclusions and Discussion}
\label{dis}

To explain the millimeter and sub-millimeter spectrum and polarization of Sagittarius A* and the recently 
observed anti-correlation between the electron temperature and the Thomson optical depth of Cygnus X-1 in the 
low-hard and intermediate states, we show that electrons need to be heated efficiently by turbulence plasma 
waves. Coulomb collisions are not effective enough to heat electrons to the temperature required by 
observations of Sagittarius A* that has a very low accretion rate. 

The critical mass accretion rate, below which a two-temperature solution may exist, is determined by the 
radiative cooling and Coulomb collisional heating processes and is almost independent of the electron heating 
by turbulence. To reproduce the observed high luminosity of X-ray binaries in the low accretion states, a high 
viscosity is required to increase the critical mass accretion rate, and the radial velocity of the accretion 
flow can be comparable to the Keplerian velocity. We show that the electron cooling for galactic X-ray binaries 
needs to be dominated by IC of an external soft photon field originated presumably from a large-scale cold slim disk. Synchrotron, synchrotron self-Comptonization, and 
bremsstrahlung processes inside the hot accretion flows are not efficient enough to cool the electrons to the observed temperature range. Assuming that the energy density of these external photons is 3 times higher than the local 
magnetic field energy density, we find that the electron temperature at the critical accretion rate is 
determined by this energy density ratio. A lower external photon energy density gives rise to a higher electron 
temperature. This high external photo energy density suggests that the radiation field may have important 
dynamical effects and may be partially responsible for the required high viscosity.

Recent observations of Cygnus X-1 show that the electron temperature in the IC model (Titarchuk 1994) decreases 
by more than a factor of 10 when the Thomson optical depth, which is proportional to the accretion rate, 
increases by two orders of magnitude in the low-hard and intermediate states. We show that very efficient 
electron heating by turbulence can reproduce such an anti-correlation. Without this extra heating process 
(besides the Coulomb collisional heating), the electron temperature is almost independent of the accretion 
rate. It appears that such an anti-correlation may also be explained by adjusting the properties of the 
external photo field (Esin et al. 1998). It is, however, not clear whether the observed dramatic changes in 
both the electron temperature and the Thomson optical depth can be reproduced in such a scenario. Detailed 
modeling of the external photo field is required to address this issue and to fit the broadband spectrum of 
Cygnus X-1 in the low-hard and intermediate states.

We have shown that a two-temperature flow can develop self-consistently at small radii for low accretion rates mostly due to the fact that electrons and protons reach a thermal equilibrium with themselves much faster than with each other through Coulomb collisions. However, at very low accretion rates, there is no guarantee that the proton distribution is Maxwellian although electrons may reach a quasi-thermal distribution due to balancing between the cooling and heating processes. Observations of the low-hard states of X-ray binaries also suggest that the electron distribution is dominated by a thermal component. 
When Coulomb collision time scales are much longer than other relevant time scales, the particle ``temperature'' should be interpreted as a measurement of the internal energy but not a thermal distribution.

With the pseudo-Newtonian potential and assumption of Keplerian azimuthal velocities, processes below $\sim 3 r_{\rm S}$ can not be described accurately by our model. These 
processes can have important observational consequences (Titarchuk \& Fiorito 2004) and may drive strong 
outflows responsible for the observed radio emission (Heinz \& Sunyaev 2003; Fender et al. 2004; Falcke et al. 
2004). By dropping these assumptions and considering the momentum conservation in the radial direction, one may 
build a more complete model to study these processes (Popham \& Gammie 1998).

In the model we have ignored the vertical structure of the disk, dynamical effects of the radiation field, 
winds and/or outflows, outwardly directed waves, and pair productions, which can be important for X-ray binaries. These may introduce a factor of a few changes in our model parameters, especially in regards to the temperature profile of protons and the transition from an one-temperature flow at large radii to the two-temperature inner region, which may have important implications on the origin of quasi-periodic oscillations. But our conclusion of the presence of efficient collision-less electron heating by 
turbulence is not affected. Electrons can also be accelerated by MHD turbulence to very high energies in 
coronas of the disk and produce the observed non-thermal high energy spectral component (Gierli\'{n}ski et al. 1999). These electrons carry much less energy than the thermal electron population considered in this paper and 
can be treated with appropriate corona models (Dermer et al. 1996; Li \& Miller 1997).

Our treatment of the electron heating by turbulence is different from that proposed in the advection dominated 
accretion flow models (Yuan et al. 2006), where the energy equation is separated into two equations, one for 
electrons and the other for ions, and it is assumed that a fixed fraction of the viscously dissipated energy is 
deposited into electrons. We point out that equation (\ref{energy}) is the more appropriate energy equation one 
should adopt and there is no reasonable justification to split it into two equations except that the coupling 
between electrons and ions through plasma waves or turbulence can be ignored. Our results and the 
MRI clearly show the opposite, i.e. turbulence plays crucial roles in an accretion 
flow. It is also difficult to understand why a fixed fraction of the viscously dissipated energy goes into 
electrons. Electron heating by plasma waves should depend on the properties of the plasma and turbulence 
(Bisnovatyi-Kogan \& Lovelace 1997; Quataert 1998).  Equation (\ref{eqte}) is perhaps more appropriate to describe the temperature evolution of 
electrons.

It is very challenging to study turbulence, one of the commonest natural phenomena. It originates from the 
non-linear nature of complex many body systems. In astrophysics, MHD turbulence is ubiquitous, and one of the 
most important effects of it is the energization of charged particles in collision-less plasmas. The 
microscopic details of this stochastic particle acceleration process are still a matter of debate. Even the 
common practice of treating MHD turbulence as a spectrum of plasma waves is not well justified, especially for 
strong turbulence. The results in this paper provide a means to study the energization of electrons and ions by 
turbulence, specifically the energy partition between the two particles, with observations. Our results suggest 
that the electron heating time by turbulence is at least 10 times shorter than that of protons at a given mean 
particle speed. Detailed modeling of the radiation spectrum and MHD simulations of the accretion flows will 
lead to more quantitative results. 

We note that the dimensionless parameter $C_1$, which characterizes the 
electron heating by turbulence in our model, decreases by a factor of $>50$ from the supermassive black hole of 
Sagittarius A* to the stellar mass black hole in Cygnus X-1. This may be related to the dramatic increase of  the turbulence viscosity described by $\beta_\nu$, which changes from 0.1 for Sagittarius A* to 3 for galactic X-ray binaries. Indeed, the product of the two changes by less than a factor of 2. 
Because $\beta_\nu > 2C_2/3$ is proportional to the energy dissipation rate of MHD turbulence, the energy 
partition between electrons and ions is described by $C_1C_2$. Our results suggest that in the 
two-temperature accretion states of black holes, this partition weakly depends on the black hole mass 
and mass accretion rate. %More detailed studies of the relevant plasma processes are required to clarify these issues.

\acknowledgments

This work was funded in part under the auspices of the U.S.\ Dept.\ of Energy, and supported 
by its contract W-7405-ENG-36 to Los Alamos National Laboratory. Some ideas of the paper were conceived while 
SL was visiting TIARA. SL thanks Dr. Ron Taam and Dr. Feng Yuan for their hospitalities and Dr. Xue-bing Wu 
and Lei Qian for useful discussions.
%This research  was partially supported by NSF grant ATM-0312344, NASA grants NAG5-12111, NAG5 11918-1 (at Stanford), NSF  grant AST-0402502 (at Arizona), and NSF grant PHY99-07949 (at KITP at UCSB). FM is very grateful to the University of Melbourne for its support (through a Miegunyah Fellowship).

\begin{figure}[bht] 
\begin{center}
\includegraphics[height=6cm]{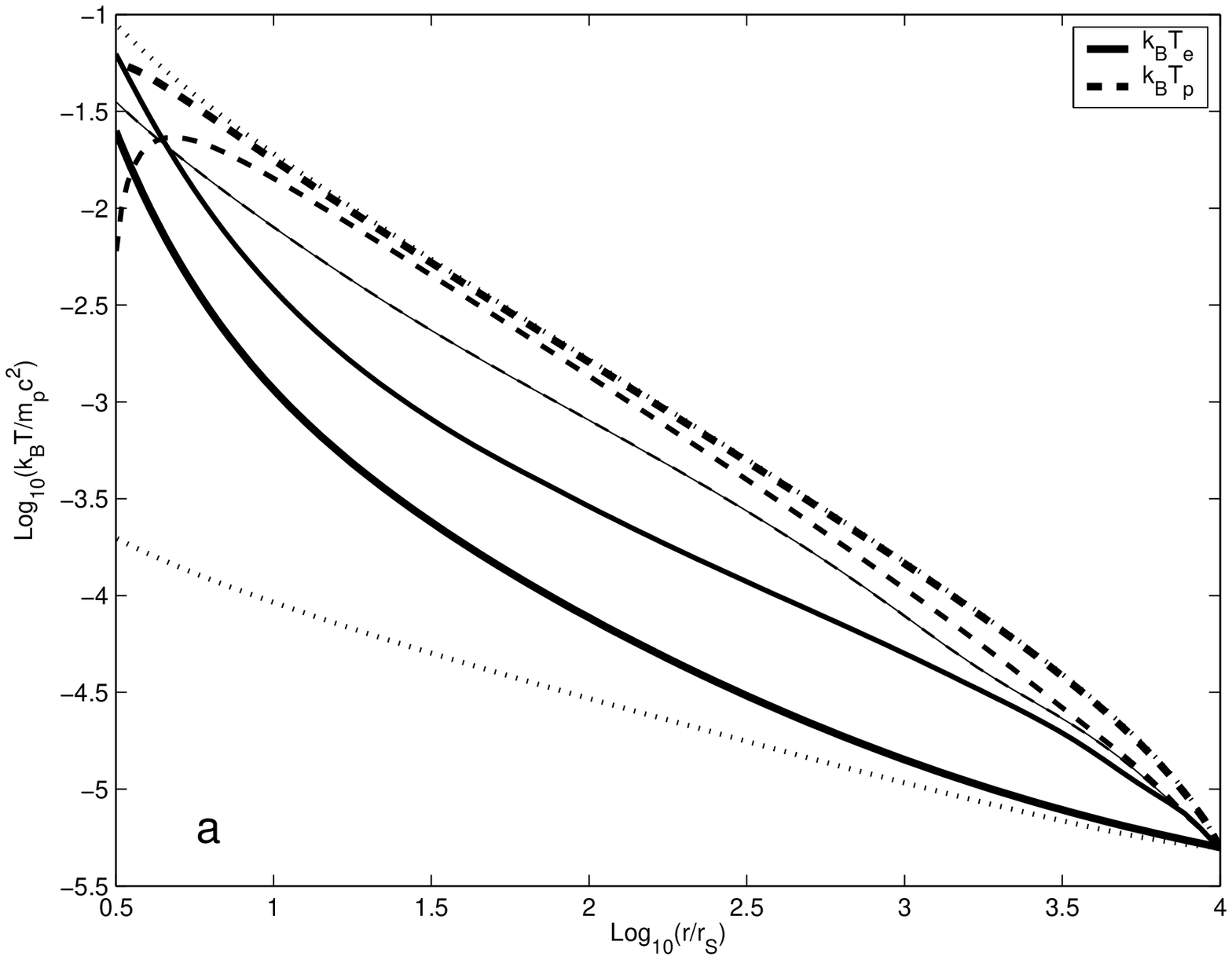}
\hspace{-0.0cm}
\includegraphics[height=6cm]{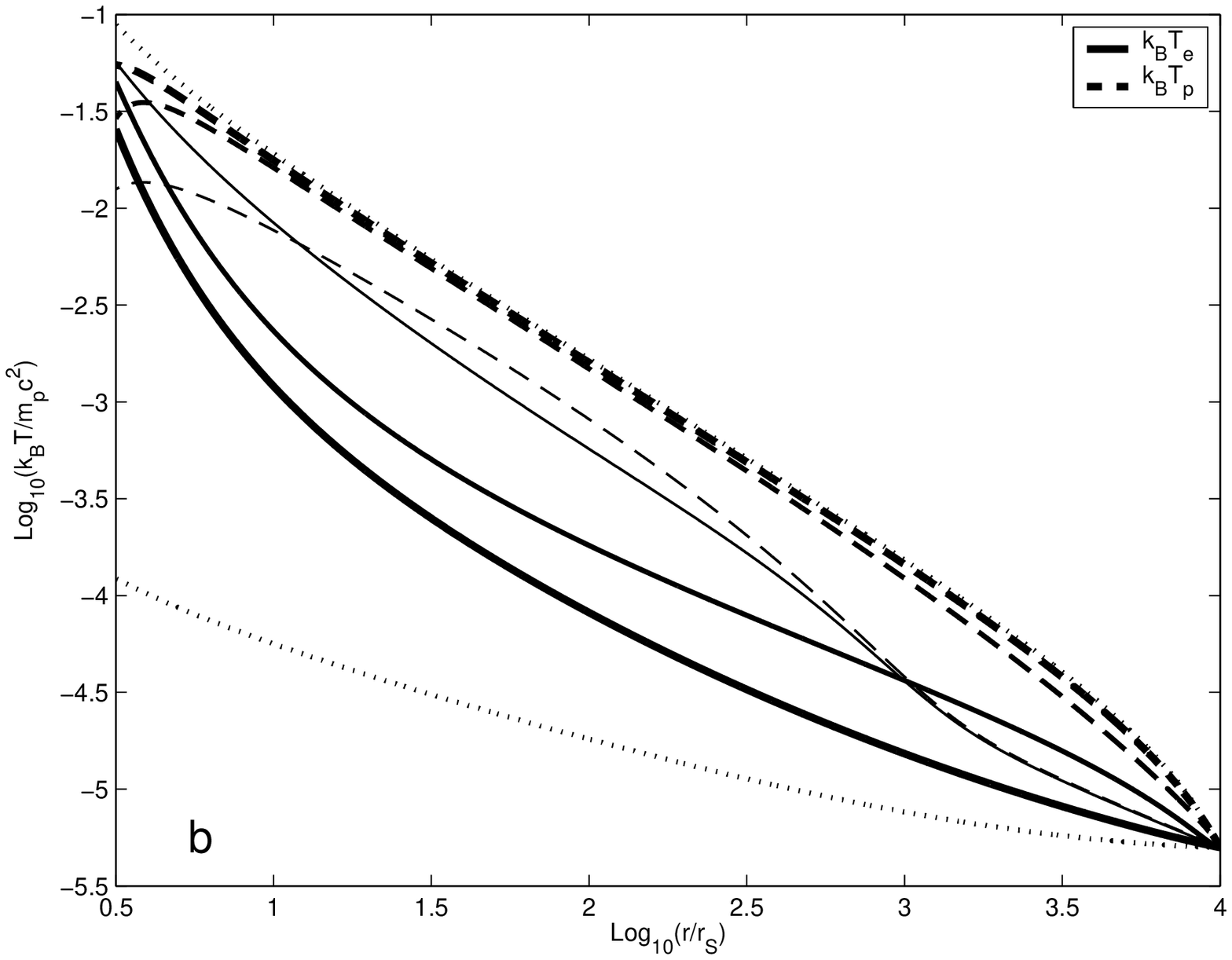}
\includegraphics[height=6cm]{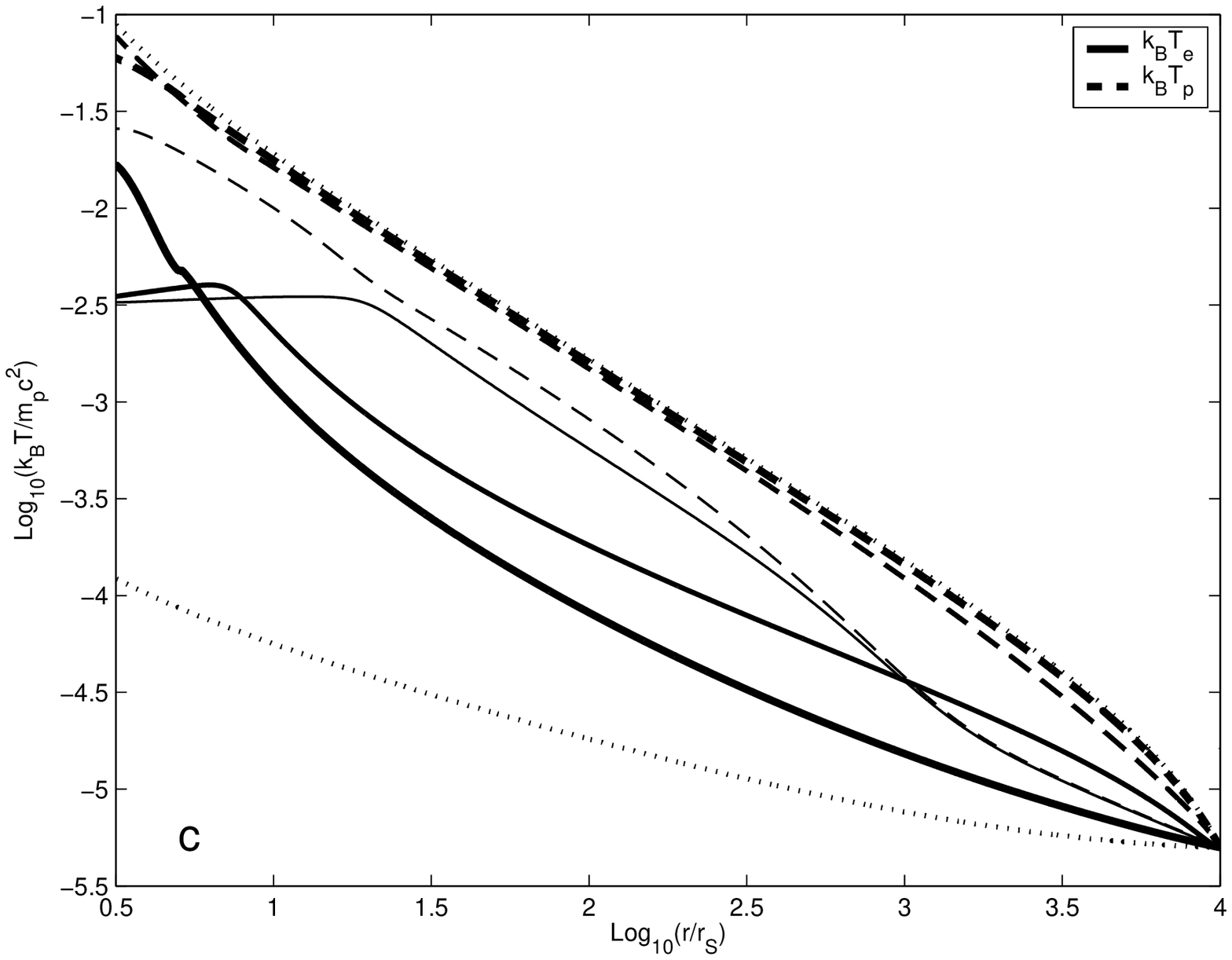}
\hspace{-0.0cm}
\includegraphics[height=6cm]{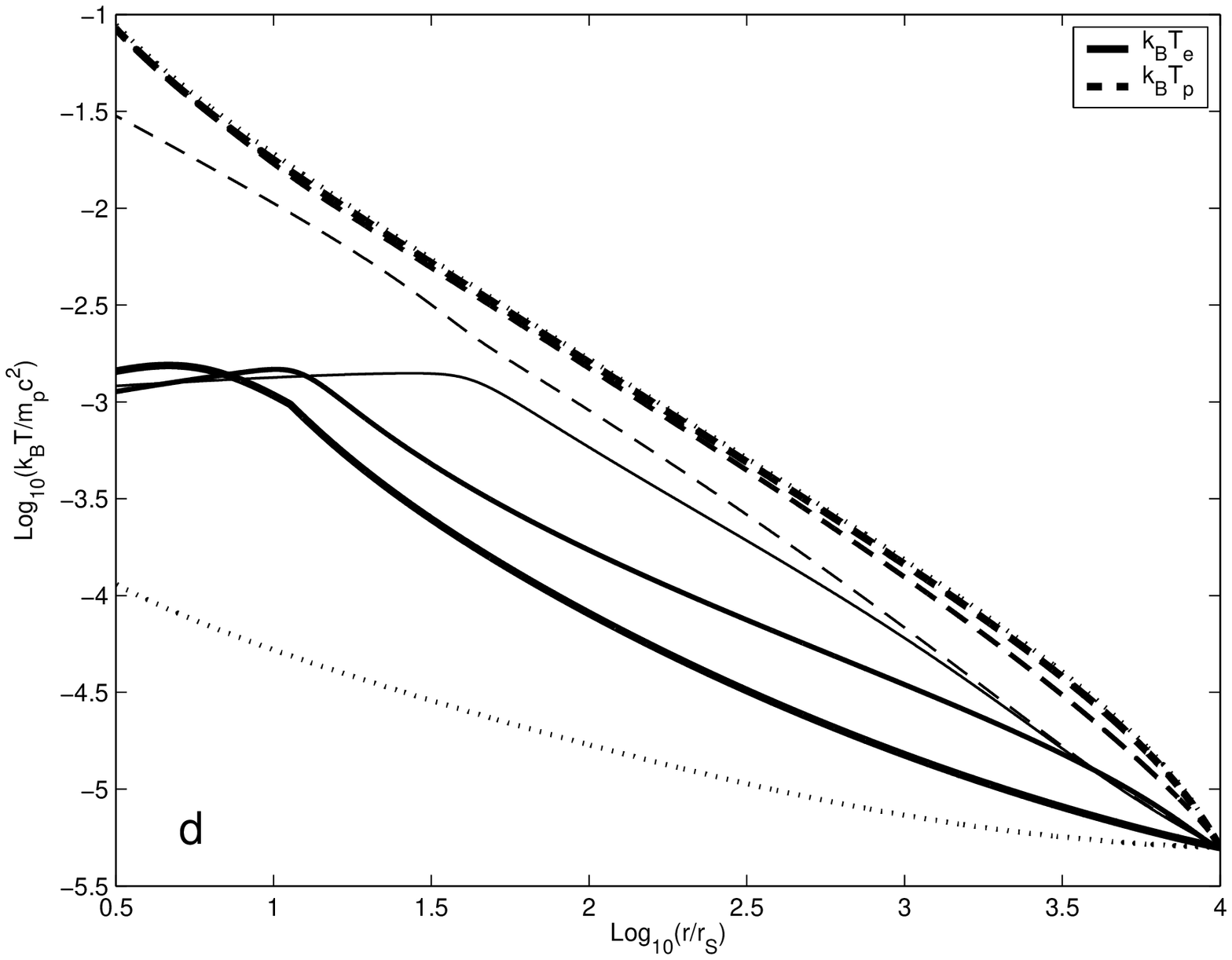}
\includegraphics[height=6cm]{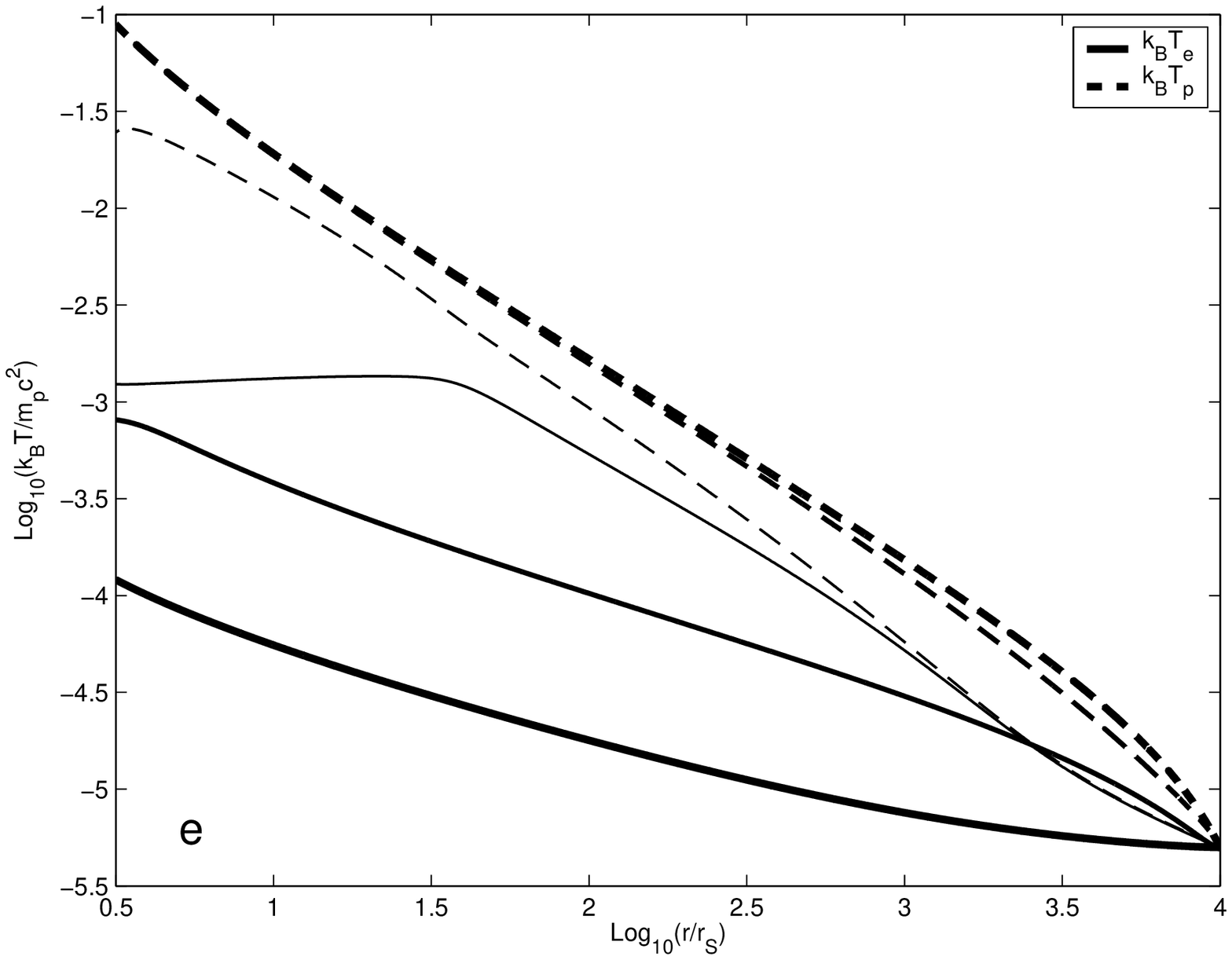}
\end{center}
\end{figure}

\begin{figure}
\caption{{\rm a:} Temperature profiles of electrons (solid) and protons (dashed) for $\dot{M} = 10^{-7}$ 
(thick), $10^{-3}$ (medium), $10$ (thin) $L_{\rm Edd}/c^2$. Cooling is not included to demonstrate the 
effects of electron heating by turbulence and Coulomb collisions. The electron and proton temperatures at 
the outer boundary are chosen so that $k_{\rm B}T_e=k_{\rm B} T_p$ is equal to $1/10$ of the gravitational 
binding energy of protons and $\varepsilon<0$. $C_1=1.5$, $\beta_\nu=0.1$, and $\beta_p=1$. The dotted lines 
have $C_1 = 15$ and $\dot{M} = 10^{-5} L_{\rm Edd}/c^2$. Note that the proton 
temperature decreases sharply toward small radii for $\dot{M} = 0.001 L_{\rm Edd}/c^2$ and the electron and 
proton temperatures are identical for $\dot{M} = 10 L_{\rm Edd}/c^2$. See text for details. {\rm b:} Similar 
to ``{\rm a}'' with the bremsstrahlung cooling included and  $\dot{M} = 2.63\times 10^{-6}$ (thick), 
$2.63\times 10^{-4}$ (medium), $2.63\times 10^{-2}$ (thin) $L_{\rm Edd}/c^2$. The disk collapses to a cold disk 
at small radii for $\dot{M}>\dot{M}_{\rm cr} = 2.63\times 10^{-2}L_{\rm Edd}/c^2(r/10^4r_{\rm S})^{-1/2}$. 
The dotted lines 
have $C_1 = 15$ and the same accretion rate as the thick lines.
{\rm
c:} Same as ``{\rm b}'' with all the cooling processes included and $M=3.4\times 10^6M_{\odot}$.  We see that 
the synchrotron and IC coolings only affect the temperature profile of electrons at small radii. {\rm d:} Same 
as ``{\rm c}'' but for $M=3.4M_\odot$. Because the synchrotron cooling time in the optically thick region 
scales as $M^{3/2}$, the electron temperature decreases with $M$ and $\dot{M}_{\rm cr} = 2.13\times 
10^{-2}L_{\rm Edd}/c^2$. {\rm e:} Same as ``{\rm d}'' but for $C_1=15$. $\dot{M}_{\rm cr} = 2.51\times 
10^{-2}L_{\rm Edd}/c^2$ }
\label{f1.eps}
\end{figure}

\begin{figure}[bht] 
\begin{center}
\includegraphics[height=6cm]{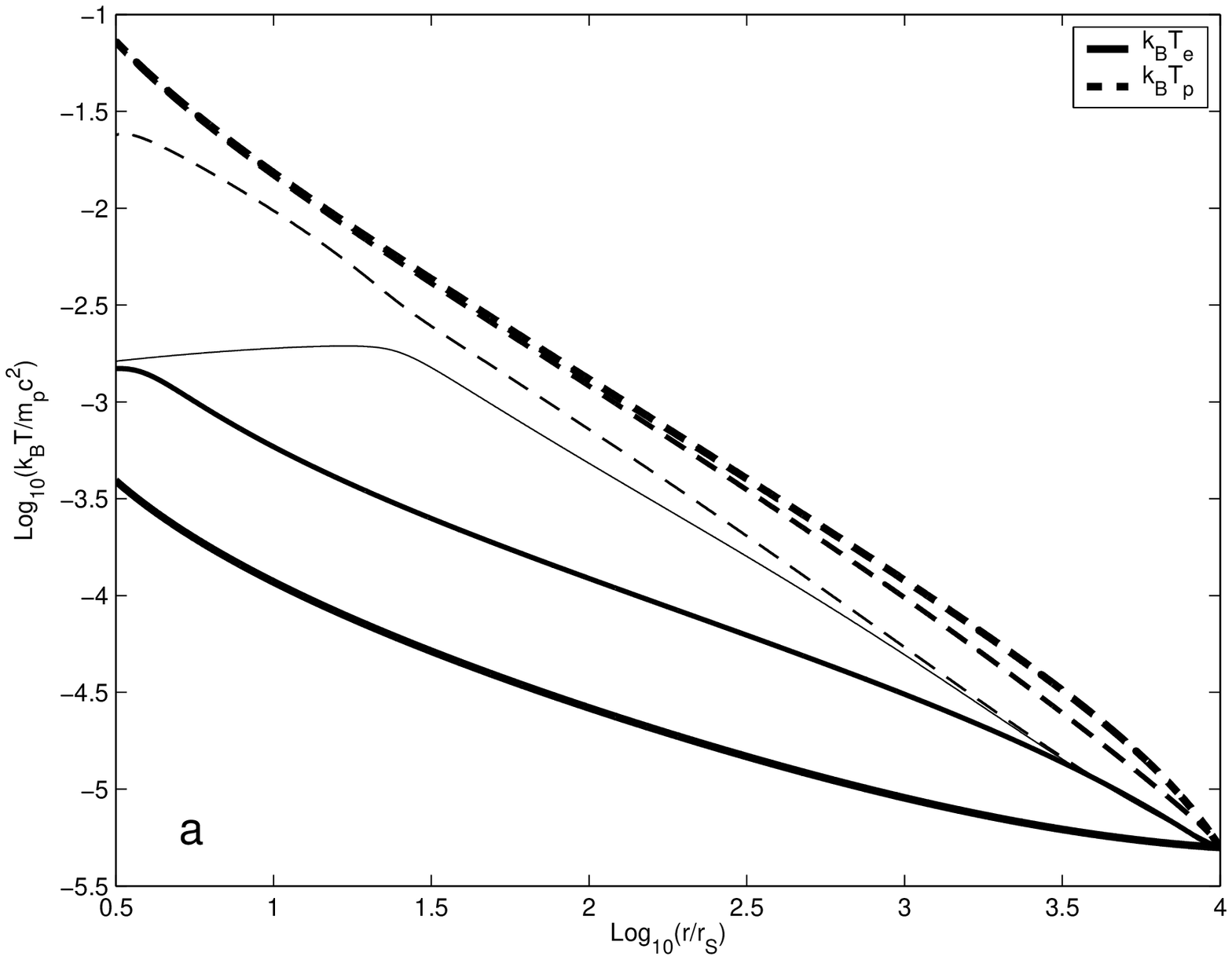}
\hspace{-0.0cm}
\includegraphics[height=6cm]{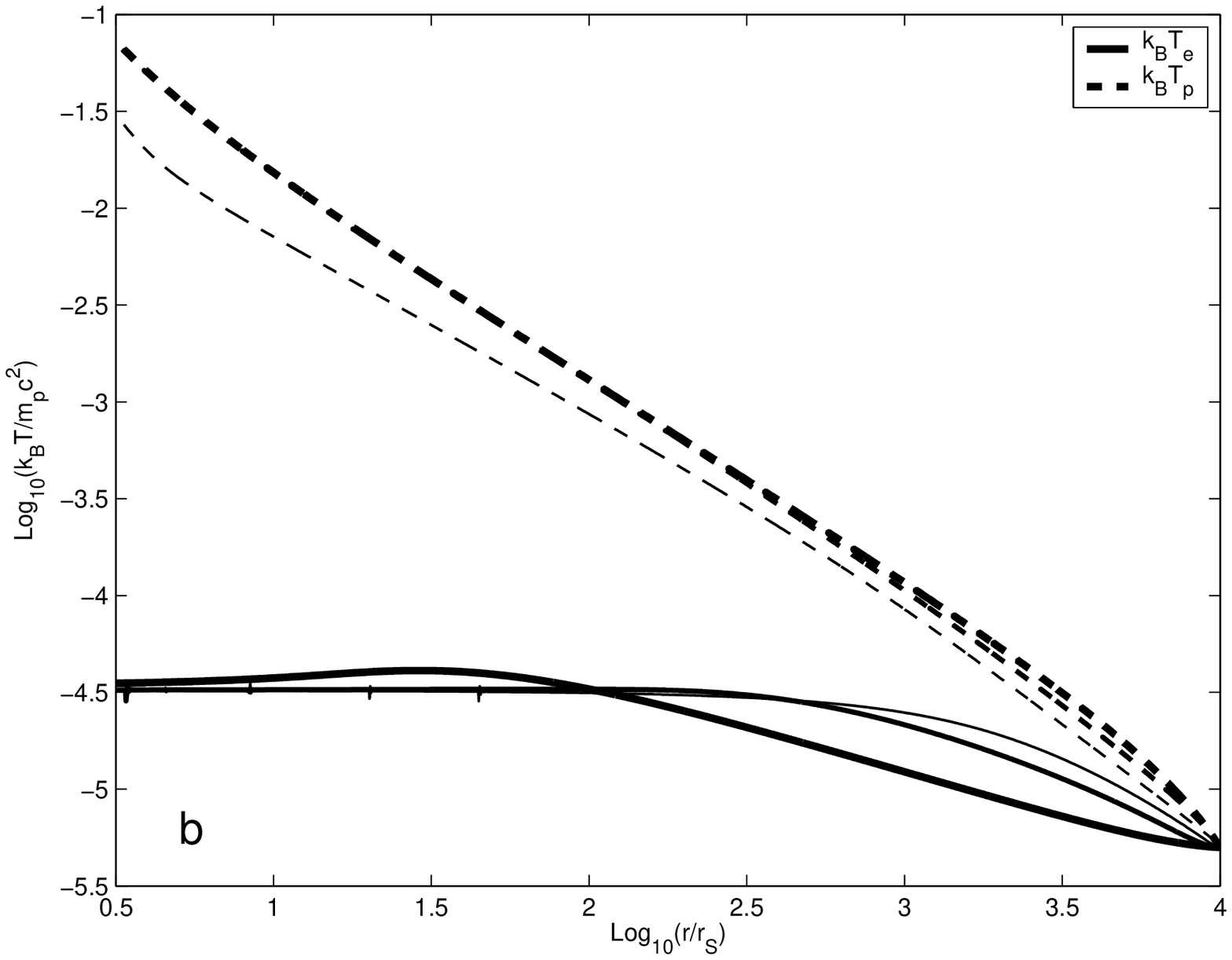}
\includegraphics[height=6cm]{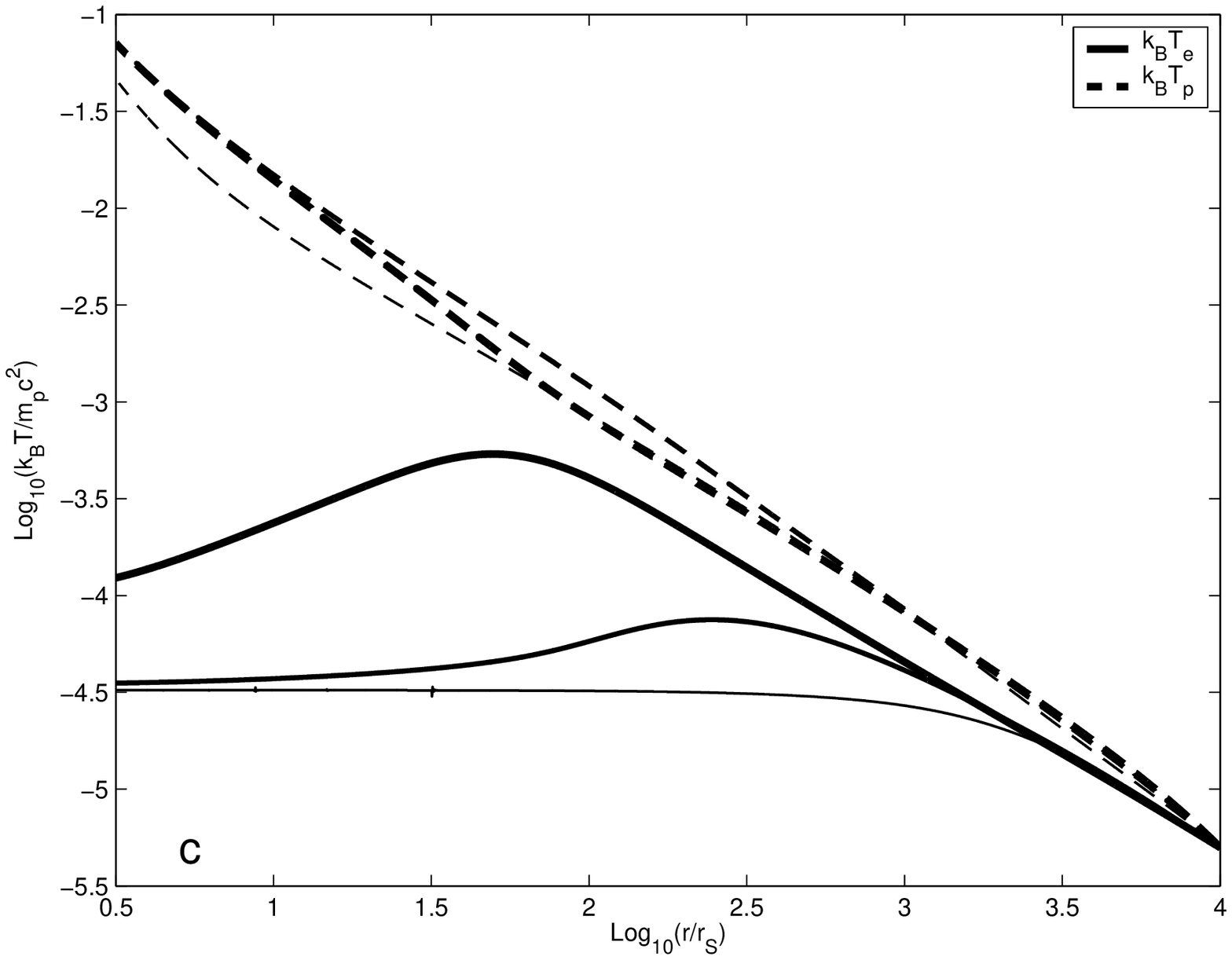}
\hspace{-0.0cm}
\includegraphics[height=6cm]{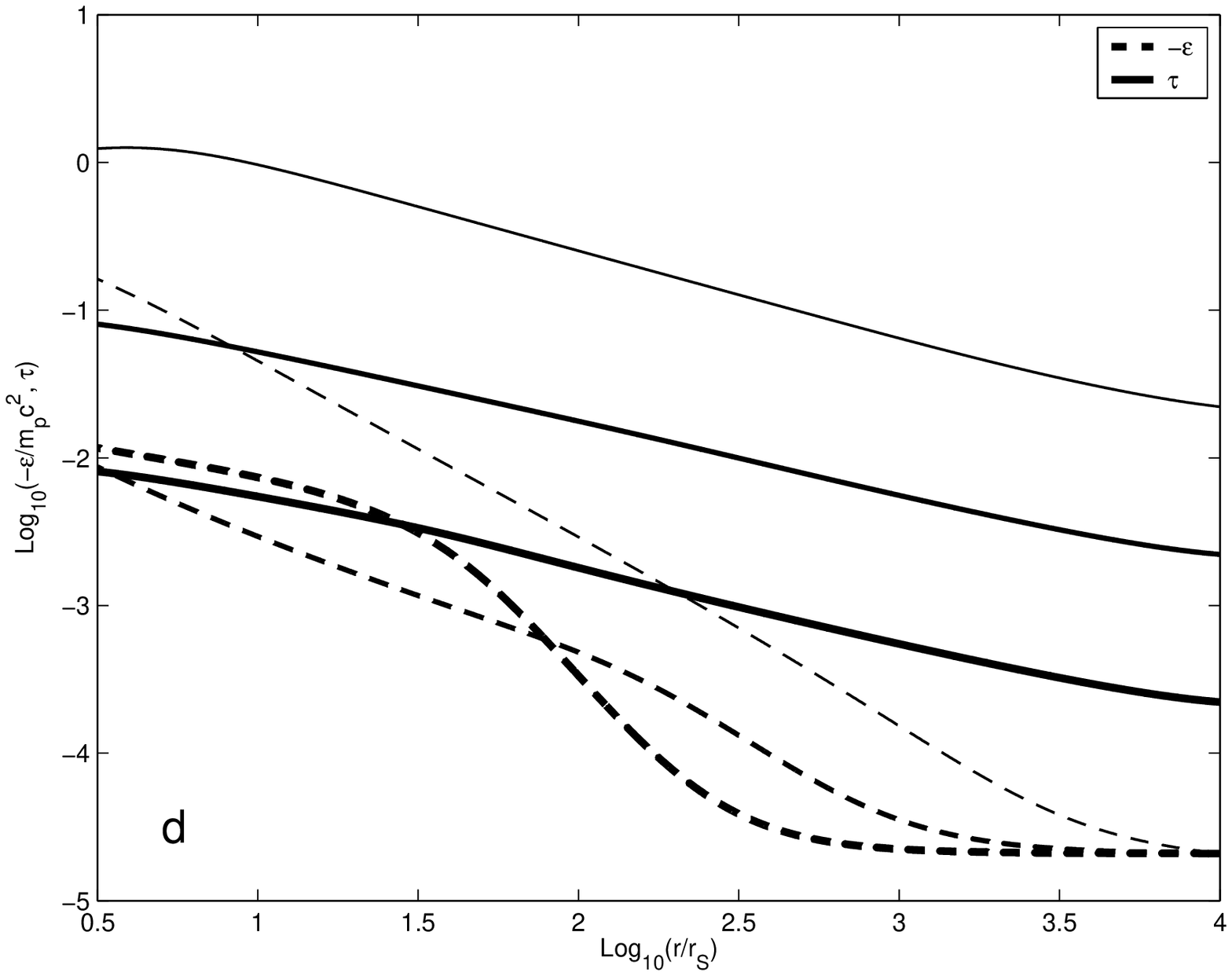}
\end{center}
\caption{
{\rm a:} Same as ``{\rm e}'' of Figure \ref{f1.eps} but for $M=34M_\odot$, $C_1=0.2$, $\beta_p = 1$, 
$\beta_\nu=3$ and $\dot{M} = 2.19\times 10^{-3}$ (thick), 0.219 (medium), 21.9 (thick) $L_{\rm Edd}/c^2$. 
In this case, $\dot{M}_{\rm cr} = 21.9 L_{\rm Edd}/c^2$. {\rm b:} Same as ``{\rm a}'' but with an external soft 
photo field, whose energy density is 3 times higher than the local magnetic field energy density. $\dot{M} = 
9.33\times 10^{-3}$ (thick), $9.33\times 10^{-2}$ (medium), 0.933 (thin) $L_{\rm Edd}/c^2$, and $\dot{M}_{\rm 
cr} = 0.933 L_{\rm Edd}/c^2$.  {\rm c:} Same as ``{\rm b}'' but for $C_1 = 0.02$, and $\dot{M} = 8.91\times 
10^{-3}$ (thick), $8.91\times 10^{-2}$ (medium), $0.891$ (thin) $L_{\rm Edd}/c^2$. $\dot{M}_{\rm cr} = 0.891 
L_{\rm Edd}/c^2$. {\rm d:} Profiles of $\varepsilon$ (dashed) and $\tau$ (solid) for the three models in  
``{\rm c}''. The thickness of the lines are the same for a given accretion rate. The luminosity of the disk 
for the thin lines is more than $9\%$  of $L_{\rm Edd}$. }
\label{f2.eps}
\end{figure}


\newpage
\begin{thebibliography}{}

\bibitem[Agol \& Krolik (2000)]{AK00} Agol, E., \& Krolik, J. H. 2000, 528, 161
\bibitem[Aitken et al. 2000]{Ait00} Aitken, D. K., Greaves, J., Chrysostomou, A., Jenness, T., Holland, W., Hough, J. H., Pierce-Price, D., \& Richer, J. 2000, ApJ, 534, L173
\bibitem[Baganoff et al. 2001] {Baganoff01} Baganoff, F. K., et al. 2001, Nature, 413, 45
%\bibitem[Baganoff et al. 2003]{Baganoff03} Baganoff, F. K., et al. 2003, ApJ, 591, 891
%\bibitem[Baganoff et al. 2005]{Baganoff05} Baganoff, F. K., et al. 2005, in preparation.
\bibitem[Balbus \& Hawley 1991]{Balbus91} Balbus, S. A., \& Hawley, J. F. 1991, ApJ, 555, L83
\bibitem[Barrio et al. (2003)]{Barrio03} Barrio, F. E., Done, C., \& Nayakshin, S. 2003, MNRAS, 342, 557
%\bibitem[Belanger et al. 2005]{Belanger06} Belanger, G., et al. 2005, ApJ, 635, 1095
%\bibitem[Belanger et al. (2006)]{Belanger05} Belanger, G., et al. 2006, \apj, submitted
\bibitem[Bisnovatyi-Kogan \& Lovelace (1997)]{BL97} Bisnovatyi-Kogan, G. S., \& Lovelace, R. V. 1997, ApJ, 486, 
L43
\bibitem[Bittner et al. (2007)]{BL07} Bittner, J., Liu, S., Fryer, C. L., \& Petrosian, V. 2007, ApJ, 661, 863
\bibitem[Blandford \& Begelman (1999)]{BB99} Blandford, R. D., \& Begelman, M. C. 1999, MNRAS, 303, L1
\bibitem[Blandford \& Eichler 1987]{Bland87} Blandford, R., \& Eichler, D. 1987, Phys. Report, 154, 1
%\bibitem[Blumenthal \& Gould 1970]{Blum70} Blumenthal, G. R., \& Gould, R. J. 1970, Rev. of Mod. Phys. 42, 237
%\bibitem[Broderick \& Loeb 2005]{Brod05} Broderick, A. E., \& Loeb, A. 2005, MNRAS, 363, 353 
%\bibitem[Bromley et al. 2001]{brom01} Bromley, B., Melia, F., \& Liu, S. 2001, ApJ, 555, L83
\bibitem[Camenzind (2005)]{Cam05} Camenzind, M. 2005, MmSAI, 76, 98
\bibitem[Dermer, Miller, \& Li 1996]{dermer96}Dermer, C. D., Miller, J. A., \& Li, H. 1996, ApJ, 456, 106
%\bibitem[Eckart et al. 2004] {Eckart04} Eckart, A., et al. 2004, A\&A, 427, 1
%\bibitem[Eckart et al. 2006] {Eckart06} Eckart, A., et al. 2006, A\&A, in press
%\bibitem[Eisenhauer et al. 2005] {Eisen05} Eisenhauer, F., et al. 2005, \apj, 628, 246
\bibitem[Esin et al. (1998)]{Esin98} Esin, A. A., Narayan, R., Cui, W., Grove, J. E., \& Zhang, S. N. 1998, ApJ, 505, 854
\bibitem[Falcke et al. (2004)]{Falcke04} Falcke, H., K\"{o}rding, E., \& Markoff, S. 2004, A\&A, 414, 895
\bibitem[Fender et al. (2004)]{Fender04} Fender, R. P., Belloni, T. M., \& Gallo, E. 2004, MNRAS, 355, 1105
%\bibitem[Fermi 1949]{Fermi49} Fermi, E. 1949, Phys. Rev. 75, 1169
%\bibitem[Galeev, Rosner, \& Vaiana 1979] {galeev} {Galeev, A. A., Rosner, R., \& 
%Vaiana, G. S., 1979}, \apj, { 229}, { 318}
\bibitem[Genzel, et al. 2003]{Genzel03} Genzel, R. et al. 2003, Nature, 425, 934
%\bibitem[Ghez et al. 2004]{Ghez04} Ghez, A. M., et al. 2004, ApJ, 601, L159
%\bibitem[Ghez et al. 2005]{Ghez05} Ghez, A. M., et al. 2005, ApJ, 635, 1087
\bibitem[Ghez et al. 2005]{Ghez05} Ghez, A. M., et al. 2005, ApJ, 620, 744
\bibitem[Gierli\'{n}ski \& Done (2002)]{GD02} Gierli\'{n}ski, M., \& Done, C. 2004, MNRAS, 329, 7  
\bibitem[Gierli\'{n}ski et al. (1999)]{Gier99} Gierli\'{n}ski, M., et al.  1999, MNRAS, 309, 496  
\bibitem[Gillessen et al. 2006]{Gil06} Gillessen S., et al. 2006, ApJ, 640, 163L
%\bibitem[Goldwurm et al. 2003]{Gold03} Goldwurm, A. et al. 2003, \apj, 584, 751
%\bibitem[Haardt \& Maraschi 1991] {hm91} {Haardt F., \& Maraschi L. 1991}, \apj, {380}, { L51}
%\bibitem[Haardt \& Maraschi 1993] {hm93} {Haardt F., \& Maraschi L. 1993}, \apj, {413}, { 507}
%\bibitem[Haardt et al 1994] {haardt94} {Haardt F., et al., 1994}, \apj, {432}, { L95}
\bibitem[Hawley \& Balbus (2002)]{HB02} Hawley, J. F., \& Balbus, S. A. 2002, ApJ, 573, 738
\bibitem[Heinz \& Sunyaev (2003)]{HS03} Heinz, S., \& Sunyaev, R. A. 2003, MNRAS, 343, L59
%\bibitem[Hornstein et al. 2006]{Horn06} Hornstein, S., et al. 2006, ApJ, in preparation 
%\bibitem[Hornstein et al. 2002]{Horn02} Hornstein, S., et al. 2002, ApJ, 577, 9 
\bibitem[Li \& Miller (1997)]{LM97} Li, H., \& Miller, J. A. 1997, 478, L67
%\bibitem[Liu \& Melia (2001)] {Liu01} Liu, S., \& Melia, F. 2001, \apj, 561, L77
\bibitem[Liu \& Melia (2002)] {Liu02} Liu, S., \& Melia, F. 2002, \apjl, 566, L77
\bibitem[Liu et al. (2006a)]{Liu06a} Liu, S., Melia, F., Petrosian, V., \& Fatuzzo, M. 2006a, \apj, 647, 1099
\bibitem[Liu et al. (2006b)]{Liu06b} Liu, S., Petrosian, V., Melia, F., \& Fryer, C. L.  2006b, \apj, 648, 1020
%\bibitem[Liu, Petrosian, \& Melia 2004] {Liu04} Liu, S., Petrosian, V., \& Melia, F. 2004, \apjl, 611, L101
\bibitem[Liu et al. (2007)]{Liu07} Liu, S., Qian, L., Wu, X.-B., Fryer, C. L., \& Li, H. 2007, ApJ, submitted, arXiv: 0705.2792.
\bibitem[Mahadevan et al. 1996]{Maha96} Mahadevan, R., Narayan, R., \& Yi, I. 1996, \apj, 465, 327
%\bibitem[Markoff et al. 2001]{Markoff01} Markoff, S., Falcke, H., Yuan, F., \& Biermann, P. L. 2001, 379, L13
%\bibitem[Melia 2006]{Melia06} Melia, F. 2006, {\rm The Galactic Supermassive Black Hole}, Princeton University Press.
%\bibitem[Melia 1992]{Melia92} Melia, F. 1992, \apjl, 387, L25
\bibitem[Melia \& Coker (1999)]{MC99} Melia, F., \& Coker, R. 1999, ApJ, 511, 750
\bibitem[Melia et al. (2000)]{Melia00} Melia, F., Liu, S., \& Coker, R. 2000, \apjl, 545, L117
\bibitem[Melia et al. (2001)]{Melia01} Melia, F., Liu, S., \& Coker, R. 2001, \apj, 553, 146
\bibitem[Meyer-Hofmeister et al. (2005)]{ML05} Meyer-Hofmeister, E., Liu, B. F., \& Meyer, F. 2005, A\&A, 432, 181 
%\bibitem[Miller et al. (1996)]{Miller96} Miller, J. A., LaRosa, T. N., \& Moore, R. L. 1996, ApJ, 
%461, 445
%\bibitem[Nayakshin \& Melia 1997a]{Nayak97a} Nayakshin, S., \& Melia, F. 1997a, \apjl, 490, L13
%\bibitem[Nayakshin \& Melia 1997b]{Nayak97b} Nayakshin, S., \& Melia, F. 1997b, \apjl, 484, L103
%\bibitem[Nayakshin, Cuadra, \& Sunyaev 2004]{Nayak04} Nayakshin, S., Cuadra, J., \&  Sunyaev, R. 2004, A\&A, 413, 173
\bibitem[Narayan \& Yi (1995)]{NY95} Narayan, R., \& Yi, I. 1995, ApJ, 452, 710 
%\bibitem[Pacholczyk 1970]{Pac70} Pacholczyk, A. G. 1970, Radio Astrophysics (San Francisco: Freeman), 86
\bibitem[Paczy\'{n}sky \& Witta (1980)]{PW80} Paczy\'{n}sky, B., \& Wiitta, P. J. 1980, A\&A, 88, 23
%\bibitem[Park \& Petrosian 1995]{Park95} Park, B. T., \& Petrosian, V. 1995, ApJ, 446, 699
%\bibitem[Parker 1979]{Parker79} Parker, E. N. 1979, Cosmical Magnetic Fields,
%Clarendon Press, Oxford
\bibitem[Pessah et al. (2006)]{Pes06} Pessah, M. E., Chan, C. K., \& Psaltis, D. 2006, PRL, 97, 221103
\bibitem[Petrosian 1981]{Petr81} Petrosian, V. 1981, \apj, 251, 727
%\bibitem[Petrosian \& Liu 2004]{Petr04} Petrosian, V., \& Liu, S. 2004, ApJ, 610, 550
%\bibitem[Petrosian \& Liu 2006]{Petr06} Petrosian, V., \& Liu, S. 2006, Conference Proceedings, submitted.
\bibitem[Popham \& Gammie (1998)]{PG98} Popham, R., \& Gammie, C. F. 1998, ApJ, 504, 419
%\bibitem[Porquet et al. 2003]{Porquet03} Porquet, D., et al. 2003, A\&A, 407, L17
%\bibitem[Qian et al. (2007)]{Q07} Qian, L. et al. 2007, ApJ, in preparation.
\bibitem[Quataert (1998)]{Qua98} Quataert, E. 1998, ApJ, 500, 978
\bibitem[Quataert \& Gruzinov (2000)]{QG00} Quataert, E., \& Gruzinov, A. 2000, ApJ, 545, 842
\bibitem[Readhead (1994)]{Read94} Readhead, A. C. S. 1994, ApJ, 426, 51
\bibitem[Rees et al. (1982)]{Ree82} Rees, M. J., Begelman, M. C., Blandford, R. D., \& Phinney, E. S. 1982, Nature, 295, 17 
\bibitem[Remillard et al. (1999)]{Rem99} Remillard, R. A., Morgan, E. H., McClintock, J. E., Bailyn, C. D., \& Orosz, J. A. 1999, ApJ, 522 397 
\bibitem[Rybicki \& Lightma (1979)]{RL79} Rybicki, G., \& Lightman, A. 1979, Radiative Processes in Astrophysics (New York: Wiley)
\bibitem[Sch\"{o}del et al. (2002)]{Sch02} Sch\"{o}del, R., et al. 2002, Nature, 419, 694
\bibitem[Shakura \& Sunyaev (1973)]{SS73} Shakura, N. I., \& Sunyaev, R. A. 1973, A\&A, 24, 337
\bibitem[Shapiro et al. (1976)]{S76} Shapiro, S. L., Lightman, A. P., \& Eardley, D. M. 1976, ApJ, 204, 187 
\bibitem[Sharma et al. (2007)]{SQHS07} Sharma, P., Quataert, E., Hammett, G. W., \& Stone, J. M. 2007, ApJ, 
submitted, astro-ph/0703572 
\bibitem[Shen et al. (2005)]{Shen05} Shen, Z. Q., Lo, K. Y., Liang, M. C., Ho, P. T. P., \& Zhao, J. H. 2005, Nature, 438, 62
%\bibitem[Schlickeiser 1984]{Sch84} Schlickeiser, R. 1984, A\&A, 136, 227
%\bibitem[Schlickeiser 1998]{Sch98} Schlickeiser, R., \& Miller, J. 1998, ApJ, 492, 352
%\bibitem[Serabyn et al. 1997]{Ser97} Serabyn, E., et al. 1997, ApJ, 490, L77   
\bibitem[Spitzer 1962]{Sp62} Spitzer, L. S. 1962, {Physics of Fully Ionized Gases}, (Dover Publications, Inc. Mineola, New York)
%\bibitem[Tagger \& Melia 2006]{Tag06} Tagger, M., \& Melia, F. 2006, ApJ, 636, L33
\bibitem[Titarchuk (1994)]{Tit94} Titarchuk, L. 1994, ApJ, 434, 570
\bibitem[Titarchuk \& Fiorito (2004)]{TF04} Titarchuk, L., \& Fiorito, R. 2004, ApJ, 612, 988
\bibitem[Wilms et al. (2006)]{Wil06} Wilms, J., Nowak, M. A., Pottschmidt, K., Pooley, G. G., \& Fritz, S. 2006, A\&A, 447, 245
\bibitem[Wolfe \& Melia (2006)]{WM06} Wolfe, B., \& Melia, F. 2006, ApJ, 638, 125
%\bibitem[Yan \& Lazarian 2004]{Yan04} Yan, H. R., \& Lazarian, A. 2004, ApJ, 614, 757
%\bibitem[Yuan et al. 2003]{Yuan03} Yuan, F., Quataert, E., \& Narayan, R., 2003, 598, 301
%\bibitem[Yuan et al. 2004]{Yuan04} Yuan, F., Quataert, E., \& Narayan, R., 2004, 606, 894
\bibitem[Yuan et al. 2006]{Yuan06} Yuan, F., Taam, R. E., Xue, Y., \& Cui, W. 2006, ApJ, 636, 46
%\bibitem[Yusef-Zadeh et al. 2006]{Yu06} Yusef-Zadeh, F., et al. 2006, ApJ, in press.
\bibitem[Zdziarski et al. (2002)]{Zd02} Zdziarski, A. A., Poutanen, J., Paciesas, W. S., \& Wen, L. Q. 2002, ApJ, 578, 357 
\bibitem[Zhang et al. (2000)]{Zhang00} Zhang, S. N., et al. 2000, Science, 287, 1239 
%\bibitem[Zhao et al. (2004)]{Zhao04} Zhao, J. H., Herrnstein, R. M., Bower, G. C., Goss, W. M.,  \& Liu, S. M. 2004, ApJ, 603, L85

\end{thebibliography}
\end{document}